\documentclass[10pt]{elsarticle}    

\usepackage{amsfonts}
\usepackage{amsmath}
\usepackage{graphicx}
\usepackage{subcaption}
\usepackage{wrapfig}
\usepackage{booktabs}

\newcommand{\bmat}[1]{\begin{bmatrix}#1\end{bmatrix}}
\newcommand{\norm}[1]{\left\lVert{#1}\right\rVert}

\newcommand{\half}{\frac{1}{2}}

\let\bbl\Bigl

\let\bbr\Bigr

\newlength{\temp}
\begin{document}

\begin{frontmatter}

\title{The Orbital Mechanics of Space Elevator Launch Systems}	
\author[Peet]{Matthew M. Peet 
\thanks{This work was supported by the National Science Foundation under grant No. 1739990.}
}
\address[Peet]{\textsc{School for the Engineering of Matter, Transport and Energy, Arizona State University, Tempe, AZ, 85298 USA.}}


\begin{abstract}
The construction of a space elevator would be an inspiring feat of planetary engineering of immense cost and risk. But would the benefit outweigh the costs and risks? What, precisely, is the purpose for building such a structure? For example, what if the space elevator could provide propellant-free (free release) orbital transfer to every planet in the solar system and beyond on a daily basis? In our view, this benefit might outweigh the costs and risks. But can a space elevator provide such a service? In this manuscript, we examine 3 tiers of space elevator launch system design and provide a detailed mathematical analysis of the orbital mechanics of spacecraft utilizing such designs. We find the limiting factor in all designs is the problem of transition to the ecliptic plane. For Tiers 1 and 2, we find that free release transfers to all the outer planets is possible, achieving velocities far beyond the ability of current Earth-based rocket technology, but with significant gaps in coverage due to planetary alignment. For Tier 3 elevators, however, we find that fast free release transfers to all planets in the solar system are possible on a daily basis. Finally, we show that Tier 2 and 3 space elevators can potentially use counterweights to perform staged slingshot maneuvers, providing a velocity multiplier which could dramatically reduce transit times to outer planets and interstellar destinations.
\end{abstract}
	
\end{frontmatter}
		
\section{Introduction}
Let us tell the story of mankind and the great arms by which it might one day reach the heavenly bodies~\cite{ferry_book,hohmann_book}.

The first chapter in this story begins with the multi-stage rocket, proposed by Konstantin Tsiolkovskii~\cite{andrews_book}, and without which space travel would be impossible. The rocket equation and its natural extension to multi-stage rockets has freed us from our terrestrial origins and given us the means to populate the Earth's Sphere of Influence (SOI) with both men and machines. From 1957 to 2020, 40 countries have launched more that 8,900 satellites and 566 humans into Earth's SOI, while 12 of those humans have traversed the largest natural object therein (the moon). And yet, no human has left the SOI of the Earth, and the number of spacecraft to depart this SOI is remarkably small (53), only 5 of which are expected to leave the solar system. Significantly, of the 5 spacecraft expected to leave the SOI of the Sun, none has achieved this feat using rockets alone, instead relying on gravitational assist maneuvers.

Beyond the SOI of the Earth, it seems man is bound by the unforgiving calculus of diminishing returns dictated by this otherwise liberating rocket equation. For example, if we were to launch a spacecraft to Alpha Centauri, with an excess velocity of 10km/s, a quick study indicates that without a gravitational assist, we would need a minimum velocity change of approximately 28 km/s (10km/s to establish a parking orbit, and 17.6 km/s to depart the SOI of the Earth and then the Sun). In the ideal case, assuming a rocket with zero structural mass and $I_{sp}=300s$, for every pound of payload, we would require 11,308 pounds of propellant. Thus, if we were to launch a small craft of only 2000 pounds to Alpha Centauri (approximately the weight of Voyager I), the smallest possible rocket (as determined by the rocket equation) would weigh more than the Eiffel tower and take 141,800 years to arrive.

And so, perhaps, there will come a time when we stop using rockets to launch towers and instead find a way to use towers to launch rockets. Perhaps we will turn the page on rockets altogether and explore a new means of space travel, starting again with Konstantin Tsiolkovskii, building a new set of rocket equations - a set of equations governed not just by gravity and mass reaction, but assisted by centripetal acceleration and the almost boundless rotational inertia of the Earth. Perhaps we will build a space elevator.


\paragraph{The Space Elevator} In simplest form, the space elevator is tower or cable, fixed to the Earth (at $r_e=6,378km$) at a base station, and rising past geosynchronous orbit ($r_g=42,164km$), with an anchor of some sort at the apex radius, $r_p$.

To understand the basic physics of a space elevator, first consider a normal Earth-orbiting satellite. The quasi-stationary motion of the orbiting satellite is dictated by the balance between gravity and centripetal acceleration (so that $\omega^2 r=\frac{F_g}{m}=\frac{\mu}{r^2}$). For a satellite to stay in orbit, then, it must have sufficient angular velocity
\[
\omega=\sqrt{\frac{\mu}{r^3}}.
\]
This angular velocity is achieved by accelerating the satellite to a linear velocity of $v=\omega r$, so that
\[
v=\omega r=r\sqrt{\frac{\mu}{r^3}}=\sqrt{\frac{\mu}{r}}.
\]
The mechanics of a space elevator are similar, except that the space elevator does not require acceleration. Instead, it is permanently attached to the surface of the Earth ($r_e =6,378 km$) and its angular velocity is then fixed by the rate of rotation of the Earth ($\omega_e=2\pi rad/day$). Thus, if the space elevator were like us (fairly small and close to the Earth), it would obviously be overcome by gravity and fall directly back to the surface (as towers sometimes do). Indeed, we should all be grateful for this - for if the Earth were not so small and its rotation were not so slow, (if we had $\omega_e^2 r_e>\frac{\mu}{r_e^2}$), then we would all fly off into space. Specifically, the critical radius of the Earth beyond which we would all be thrown into space is $42,164km$.

However, unlike humans or satellites, the length of a space elevator spans the distance from the surface of the Earth ($r_e =6,378 km$) to its apex (at $r_p>42,164 km$). Thus, while the lower sections ($r<42,164 km$) of the elevator tend to fall back to Earth, the sections higher than $42,164km$ tend to fly off into space. Assuming, then, that the space elevator does not tear itself apart, the higher sections will prevent the lower sections from falling back to Earth. Meanwhile, the lower sections (and an attachment at the base) prevent the higher sections from flying off into space.

And yet, what role can the space elevator play when, the time and the technology being right, humanity decides that flying off into space is more desirable than the terrestrial alternative? Perhaps, at that time, we might wish for a planet of radius greater than $42,164km$, so that leaving our terrestrial origin would involve nothing more than stepping the front door. Can the space elevator artificially expand the radius of the Earth to the point where centripetal acceleration is stronger than gravity? Beyond this point, are rockets even necessary to realize our interplanetary and interstellar ambitions? In this manuscript, we will carefully examine all these questions, starting with a tiered description of a space elevator launch system, proceeding to the derivation of a space elevator variant of the rocket equation, considering the problem of launch into the ecliptic plane, determining the excess velocity envelopes at exit of the Earth SOI, and calculating the minimum TOF to various heavenly bodies as a function of the relative alignment of the planets.

\paragraph{Contribution of the Manuscript}
The concept of a space elevators has been around since Tsiolkovskii originally proposed it in 1895~\cite{tsiolkovskii_1895} and various iterations of the concept have been developed in, e.g.~\cite{artsutanov_1960,isaacs_1966,pearson_1975}. An entertaining introduction to the history and various re-inventions of the space elevator can be found in~\cite{clarke_1981}.

Of course the first thing to understand about the space elevator is that there are doubts about whether it can be built using existing materials. Specifically, the question centers on whether there exists a material (such as single crystal graphene) with sufficient tensile strength so that the elevator does not tear itself apart at the point of maximum stress, which occurs at radius $r_g=42,164$. This issue has been discussed in great depth and hence we ignore it, referring instead to a selection of articles on the subject~\cite{pugno_2007,aslanov_2013,edwards_2000,aravind_2007,popescu_2018}. Another question frequently raised is the dynamic stability of the structure itself as climbing and sliding spacecraft move along the length of the tower. Here the doubts seem to be less acute or can be mitigated with counterweights~\cite{mcinnes_2006}, and we again refer to some more carefully selected references such as~\cite{cohen_2009,li_2020,jung_2014,pugno_2009,perek_2008,sadov_2011,steindl_2005}.

While there has been a great deal of effort looking at whether one can build a space elevator, it seems that considerably less work has been done to answer the question of what one can do with such an elevator~\cite{swan_2006}. Indeed, there does not even seem to be a consensus on how a space elevator launch system would work. Specifically, there are two sources of velocity which can be exploited to accelerate a spacecraft out of the Earth's SOI. The first is the tangential velocity of the apex anchor itself, $v_t=\omega_e r_p$. If the elevator is long enough, this velocity by itself is sufficient to escape the Earth SOI (We call this design a Tier 0 elevator), and most of the few papers treating the problem of interplanetary travel from a space elevator launch system only consider this source of velocity~\cite{chobotov_2004,gao_2016,torla_2019}. However, there is a second source of velocity, wherein a sliding spacecraft is placed on the leading edge of the space elevator and allowed to accelerate freely under in the influence of centripetal acceleration to final radial velocity $v_r$ (a Tier 1 elevator). This second source of velocity has been examined in~\cite{knudsen_thesis,mcinnes_2006,pearson_1997}, although it has never been applied to the problem of interplanetary mission design. Indeed, with the possible exceptions of~\cite{chobotov_2004,torla_2019}, it seems that no detailed examination of the orbital mechanics of any space elevator launch system has ever been made. The goal, then, of this paper, is to make such a study - providing a rigorous mathematical basis, accounting for both radial and tangential velocity, detailing the required lengths for interplanetary missions, solving the problem of transfer to the ecliptic, cataloguing the excess velocity envelopes, determining minimum  to the outer planets and beyond, and finally describing and quantifying a space elevator slingshot maneuver enabled by counterweights.

\paragraph{Organization and Methods}
We will begin our study of space elevator launch systems by defining precisely what we think are the necessary requirements for a space elevator launch system. We take a tiered approach to this problem, first defining a minimal space elevator consisting of a simple cable extending to an apex radius (Tier 1). However, as we will show in Section~\ref{sec:thc=0}, such a design is inefficient, resulting in reduced launch velocities and infrequent launch windows. Our second elevator design includes an apex ramp (Tier 2). We will show in Section~\ref{sec:escape} that this ramp significantly improves launch velocities but does not result in a significant increase in frequency of launch opportunities. Our third space elevator design allows for slow rotation of the apex ramp (Tier 3). This rotation allows for daily free release transfers to all planets in the solar system, as shown in Sections~\ref{sec:ecliptic} and~\ref{sec:lambert}.

Note that we do not consider elevator-like launch systems such as sky-hooks which are not fixed to the Earth by a base anchor. The reason is simple. The massive kinetic energies imparted to the departing spacecraft must come from somewhere. For launch systems which are not fixed at the Earth, this energy comes from the energy of the orbit of the launcher. Thus, this orbital energy must be replenished using rockets, which obviates the motivation for designing a space elevator in the first place. Indeed, if the space elevator is not fixed at the base anchor, every launch will decrease the angular velocity of the center of mass of the tether, causing it to drag on the ground and drift to the west. For a space elevator which is fixed at the base station, however, any minute difference in rotation rate between the Earth and the space elevator will create a reaction force which stabilizes the elevator about the angular velocity of the Earth.

Having defined our tiered space elevator launch system, we begin our work in Section~\ref{sec:velocity}, where we perform some necessary coordinate rotations to obtain launch velocities in the Geocentric Celestial Reference Frame (GCRF). These results are not very interesting, per se, but rather defines certain velocity vectors which will enable the analysis in the rest of the manuscript.

Based on the velocity vectors produced in Section~\ref{sec:velocity}, we then state the most serious constraint facing space elevator launch systems - transfer to the ecliptic plane. Specifically, the space elevator lies in the equatorial plane, but planets lie in the ecliptic plane. The question, then, is how to manage this plane change without rockets, using only centripetal acceleration and gravity. Specifically, Section~\ref{sec:ecliptic_motivation} defines the constraints on the velocity vector at exit from the Earth's SOI which must be satisfied in all subsequent sections.

Having formulated the basic velocities and constraints involved in free release transfer from a space elevator, in Section~\ref{sec:thc=0} we examine the first tier of space elevator - determining minimum lengths, calculating required release times, examining Hohmann transfer options and, finally, determining launch windows for free release transfer to the planets and beyond. The second tier of space elevator is studied in Section~\ref{sec:escape}, where we show that use of an apex ramp significantly decreases the required length of the space elevator, but does not significantly improve the availability of launch windows.

In Section~\ref{sec:ecliptic} we then examine the third tier of elevator and show that rotation of the apex ramp allows for ecliptic transfer at any time of day. We provide a Newton-Raphson algorithm for computing the associated rotation angles, and display the resulting excess velocity envelopes which are then used in Section~\ref{sec:lambert} to calculate the minimum  to each planet as a function of day of synodic year. Finally, in Section~\ref{sec:trebuchet}, we examine the mathematics of space elevator slingshot maneuvers using a secondary apex ramp for launch of staged counterweights.

Throughout this paper, we emphasize simple mathematical relationships and analytic formulae, eschewing numerical methods whenever possible. While computational analysis is certainly a valuable tool, and indispensable in applied mission design, the field of astrodynamics has become overly fond of such algorithms - at the expense of the intuition and inspiration provided by the fundamental principles and mathematical relations which define the natural world.
To this end, as far as possible, all of our analysis will be \textit{analytic} - i.e.  we try not to rely too heavily on numerical approximations or optimization algorithms. 

And now, the stage being set, let us begin our story, of orbits and elevators, a tale which has before been sung to the rhythm of rockets - a tale that begins and ends with Konstantin Tsiolkovskii.

\section{Frames of Reference}
For reference, we include here a list of coordinate systems used in this paper, defining the origin and principle axes. Note that this is included only for reference and coordinate systems are also defined when they are first used.

\paragraph{Perifocal Coordinate System (PQW)} The origin is Earth center. $\hat x_{PQW}=\hat p$ is directed to the periapse of the orbit. $\hat z_{PQW}=\hat w$ is directed orbit-normal (aligns with angular momentum vector). $\hat y_{PQW}=\hat q$ is obtained from the right hand rule.

\paragraph{Space Elevator Inertial (SEI)} The origin is Earth center. $\hat x_{SEI}$ is directed to the apex anchor position. $\hat z_{SEI}$ is directed to the terrestrial north pole. $\hat y_{SEI}$ is obtained from the right hand rule.

\paragraph{Earth Centered Earth Fixed (ECEF)} The origin is Earth center. $\hat x_{ECEF}$ is directed to prime meridian. $\hat z_{ECEF}$ is directed to the terrestrial north pole. $\hat y_{ECEF}$ is obtained from the right hand rule.

\paragraph{Earth Centered Inertial (ECI)} The origin is Earth center. $\hat x_{ECI}$ is directed to the First Point of Ares (FPOA). $\hat z_{ECI}$ is directed to the terrestrial north pole. $\hat y_{ECI}$ is obtained from the right hand rule.

\paragraph{Geocentric Celestial Reference Frame (GCRF)} The origin is Earth center. $\hat x_{GCRF}=\hat x_{\epsilon}$ is directed to the First Point of Ares (FPOA). $\hat z_{GCRF}=\hat z_{\epsilon}$ is directed to the celestial north pole. $\hat y_{GCRF}$ is obtained from the right hand rule.

\paragraph{Barycentric Celestial Coordinate System (BCRS)} The origin is the barycenter of the solar system, which we take as Sun center. $\hat x_{BCRS}=\hat x_{\epsilon}$ is directed to the First Point of Ares (FPOA). $\hat z_{BCRS}=\hat z_{\epsilon}$ is directed to the celestial north pole. $\hat y_{BCRS}$ is obtained from the right hand rule.

\section{Space Elevator Launch Configuration}
\begin{figure}[!]
  \centering
\includegraphics[width=\textwidth]{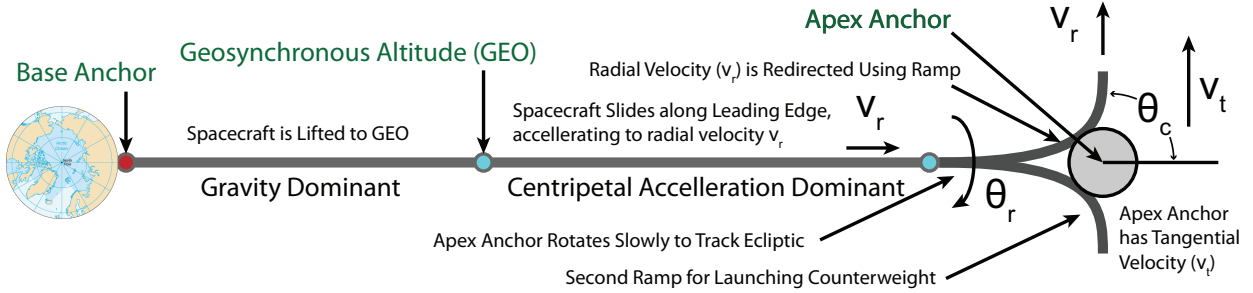}
  \caption{Baseline Space Launch Configuration. Distances and sizes are not to scale.}\label{fig:design}
\end{figure}

\begin{figure}[ht]
    \begin{subfigure}[t]{0.45\textwidth}
        \centering
\includegraphics[width=\linewidth]{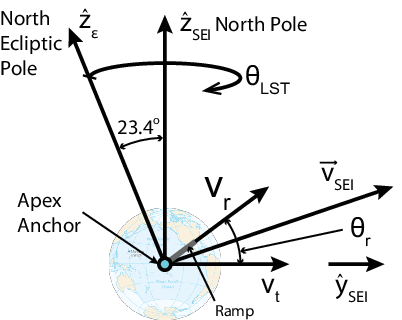}
        \caption{Illustration of launch geometry from directly above the space elevator (Equatorial view) for $\theta_c=90^\circ$. $\hat x$ points out of the page in the SEI and perifocal coordinate systems. $\hat z_{\epsilon}$ is the $\hat z$ unit vector in the GCRF and BCRS coordinate systems. $\hat z_{SEI}$ is the $\hat z$ unit vector in the SEI coordinate system. $\hat z_{\epsilon}$ is rotated about $\hat z$ by the local sidereal time, $\theta_{LST}.$ Tangential velocity of the apex anchor itself is depicted as $v_t$. Rotation of the launch ramp is depicted by $\theta_r$. Because $\theta_c=90^\circ$, the outward radial velocity, $v_r$ is initially redirected along the $\hat y_{SEI}$ unit vector. Rotation of the launch ramp by amount $\theta_r$ redirects this component in the $\hat z_{SEI}$ direction. Combining these terms yields $\vec v_{SEI}$. Ignoring turning angle, this shows that even if $\theta_r=0^\circ$, the velocity vector $\vec v_{SEI}\cdot \hat z_{\epsilon}=0$ twice a day, corresponding to $\theta_{LST}=90^\circ,\, 270^\circ$ (Section~\ref{sec:escape}). In addition, for any $\theta_{LST}$, the angle $\theta_r$ can be used to enforce $\vec v_{SEI}\cdot \hat z_{\epsilon}=0$ (Section~\ref{sec:ecliptic}).}
\label{fig:SEI}
    \end{subfigure}\hfil
    \begin{subfigure}[t]{0.45\textwidth}
        \centering
\includegraphics[width=.8\linewidth]{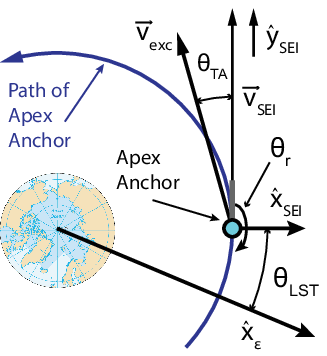}
        \caption{Illustration of launch geometry from north pole for $\theta_c=90^\circ$. The First Point of Ares (FPOA) is denoted $\hat x_{\epsilon}$ and is the $\hat x$ unit vector in the GCRF and BCRS coordinate systems. $\hat x_{SEI}$ is the $\hat x$ unit vector in the SEI and perifocal coordinate systems. These unit vectors differ by amount  $\theta_{LST}$, which is the local sidereal time. $\hat y_{SEI}$ is the $\hat y$ unit vector in the SEI coordinate system. Since $\theta_c=90^\circ$, the radial outward component of velocity, $v_r$ is rotated into the $\hat y_{SEI}$ direction. This radial component is then rotated by amount $\theta_r$ about the $\hat x_{SEI}$ axis. This radial component is then combined with the tangential velocity of the apex anchor itself to yield $\vec v_{SEI}$, which lies in the $\hat y_{SEI}-\hat z_{SEI}$ plane. The turning angle, $\theta_{TA}$ then rotates $\vec v_{SEI}$ about the orbit-normal $\hat z_{PQW}$ axis (not depicted) to obtain the excess velocity vector, $\vec{v}_{exc}$ at the exit of Earth's SOI.}
\label{fig:SEI_polar}
    \end{subfigure}
    \caption{Illustrations of the Space Elevator at Free Release.}\label{fig:elevator_figs} \vspace{-5mm}
\end{figure}

In this paper, we consider tiered modifications to the standard space elevator design. We define the standard space elevator design, as proposed by Tsiolkovskii and augmented with a sliding surface in~\cite{pearson_1975}, as a Tier 1 Space Elevator. Note that without the sliding surface introduced in~\cite{pearson_1975} to produce radial velocity (Tier 0), targeting of free release transfers to the planets becomes essentially impossible, as discussed in Sections~\ref{sec:rocket} and~\ref{sec:escape}. Beyond the sliding surface, we propose the following modifications to the Tier 1 space elevator design.
\begin{enumerate}
  \item \textbf{[Tier 2]} An apex ramp, located at the apex anchor, for redirection of radial velocity in the tangential direction. The slope of this ramp is parameterized by the angle $\theta_c \in [0^\circ,90^\circ]$. First used in Section~\ref{sec:escape}.
  \item \textbf{[Tier 3]} A pivot point for slow rotation of the apex ramp used to track the ecliptic plane. The angle of rotation is parameterized as $\theta_r \in[-90^\circ,90^\circ]$. First used in Section~\ref{sec:ecliptic}.
  \item \textbf{[Tier 3 with Slingshot]} A second apex ramp, collocated with and oriented $180^\circ$ from the primary ramp, for use in launching counterweights and performing slingshot maneuvers. First used in Section~\ref{sec:trebuchet}.
\end{enumerate}
The overall design and modifications are depicted in Figures~\ref{fig:design},~\ref{fig:SEI}, and~\ref{fig:SEI_polar}. These three modifications are added to the analysis of the manuscript incrementally, so Tier 1 elevators are studied in Section~\ref{sec:thc=0}, Tier 2 elevators are studied in Section~\ref{sec:escape}, Tier 3 elevators are studied in Sections~\ref{sec:ecliptic} and~\ref{sec:lambert}, and slingshot maneuvers using the second ramp are introduced in Section~\ref{sec:trebuchet}. Let us now examine the motivation for each of these modifications.

First, the apex ramp, oriented at an angle $\theta_c$ (which cannot be modified once constructed), allows us to utilize radial velocity more efficiently by redirecting it in the direction of the motion of the apex anchor itself. Structurally, the ramp is similar to a slide in a playground, but since the radial velocity can potentially reach 10km/s at the apex, centrifugal forces for spacecraft on this ramp can be substantial, depending on the radius of curvature. Specifically, if $r_c$ is the radius of curvature and $v_r$ is the radial velocity, the centripetal acceleration experienced by the spacecraft on the ramp will be
\[
a_c:=\omega^2 r_c=\frac{v_r^2}{r_c}.
\]
For example, if the radius of curvature is 1000km, and $v_r=10km/s$, the acceleration will be 100g's for around 100 seconds. Since no human has withstood more than 42g's, this would not be survivable without some form of pressurized survival couch~\cite{seedhouse_2012}. Force on the ramp structure itself, however, is not particularly problematic, given other assumptions on strength of materials.

Second, a pivot point is included for relatively slow rotation of the apex ramp so as to allow the apex ramp to track the ecliptic, as described in detail in Section~\ref{sec:ecliptic}. Assuming continuous tracking, the period of this rotation is 1 Earth day. As will be discussed in Section~\ref{sec:ecliptic_motivation}, the problem of transfer to the ecliptic severely limits the launch windows available for free release transfer to the planets - resulting in gaps of up to 65 years in the case of Neptune. The apex ramp rotation, by contrast, allows for daily free release transfers - significantly increasing the utility of the space elevator. By way of further justification, this part of the design seems less onerous than some proposals in, e.g.~\cite{knudsen_2016,edwards_2000}. Furthermore, the energy requirements for performing the rotation can be obviated through the use of large flywheels to store rotational inertia. Note in addition, that only the ramp need be rotated, and not the entire apex anchor.

Finally, we include in our design a secondary apex ramp, in symmetry with the primary apex ramp about the apex anchor. The purpose of this ramp is two-fold. First, recall that all net energy added to the spacecraft comes from rotation of the Earth. This energy must be transmitted through the length of the space elevator. The rapid accelerations produced by the primary ramp therefore will cause oscillations of the apex anchor and large transient strains in the elevator. However, if counterweights are launched in tandem with the spacecraft, this eliminates the large transient strains introduced by the apex ramp. In addition, as described in Section~\ref{sec:trebuchet}, a counterweight launched in tandem with and tethered to a spacecraft can be used to create a slingshot effect, similar in operation to a gravity assist, which can more than triple the redirected radial velocity. As a practical note, craft destined for the secondary ramp could potentially be suspended on the trailing edge of the elevator by attachment to a runner sliding on the leading edge, similar to a zip-line. This would make transition to the secondary ramp at apex significantly simpler.

Having described our space elevator design, we note that the modifications to the standard configuration are not chosen arbitrarily, but rather are, in our view, the minimal modifications necessary for creation of a space elevator capable of delivering on the promise of daily propellant-free launches to the outer planets and beyond - a benefit which may outweight the costs and risks of construction. Specifically, without the apex ramp and apex rotations, prograde transition to the ecliptic plane occurs only twice a day, and the outgoing velocity vector with respect the Earth is permanently fixed. These geometric constraints make successful propellant-free transfer orbits infrequent - See Subsection~\ref{subsec:T1_nonhohmann}.

\section{The Space Elevator Velocity Equation}\label{sec:rocket}
In this section, we derive an expression for radial velocity of a spacecraft sliding on the leading edge of the space elevator under the influence of both gravitational and centripetal acceleration. The use of a sliding surface to produce radial velocity was originally proposed in~\cite{pearson_1975} and the magnitude of this radial velocity is one of three variables we can control (degrees of freedom) to obtain a desired excess velocity vector at exit from the Earth's SOI, the other two being $\theta_r$ (Section~\ref{sec:ecliptic}) and release time, $\theta_{LST}$.

To begin, we recall that, as depicted in Figure~\ref{fig:design}, there are two balancing sources of acceleration in a space elevator. The first is gravitational and the second is centripetal. The critical radius beyond which centripetal acceleration dominates is that of geosynchronous orbit (GEO) (denoted $r_{g}$). Now consider what happens when a spacecraft is allowed to slide freely along the leading edge of space elevator, starting at some radius, $r_0>r_{g}$.

For this derivation, we use an approach inspired by Tsiolksovskii's rocket equation, wherein integration is performed with respect to expenditure of propellant mass, which represents the variable of potential energy. In our approach, however, the source of potential energy is not mass but rather radius, $r$, which is defined as the distance from the center of the Earth to the location of the sliding spacecraft on the space elevator. In this framework, then, we define radial acceleration $a(r)$ and radial velocity $v(r)$ to be functions of radial distance $r$ and integrate $a(r)$ over the length of the sliding surface. Specifically, if $v(r)$ is the radial speed of the spacecraft along the length of the elevator and $a(r)$ is the rate of change of $v(r)$, then
\[
a(r)=\frac{dv}{dt}=\frac{dv}{dr}\frac{dr}{dt}=v(r)\frac{dv(r)}{dr}.
\]
Now, combining acceleration due to gravity and centripetal acceleration, we obtain an expression for the left-hand side of the expression as
\[
a(r)=-\frac{\mu_e}{r^2}+\omega_e^2 r
\]
where $\mu_e$ is the gravitational constant of the Earth and $\omega_e$ is the rotation rate of the Earth.
Combining these two expressions, we obtain
\[
v dv=\left(-\frac{\mu}{r^2}+\omega_e^2 r\right) dr.
\]
Integrating both sides of the equation, we obtain
\[
\half v^2\vert_{v_0}^{v_f}=\frac{\mu}{r}+\frac{\omega_e^2}{2} r^2\vert_{r_0}^{r_f}
\]
where $v_0$ is the velocity of the spacecraft at radius $r_0$ and $v_f$ is the velocity of the spacecraft at $r_f$.
Solving for the final velocity, $v_f$ at radius $r_f$, we obtain
\begin{equation}
v(r_f,r_0)=\sqrt{v_0^2 + \frac{2\mu}{r_f}-\frac{2\mu}{r_0}+\omega_e^2 \left(r_f^2-r_0^2\right)}.\label{eqn:vf_1}
\end{equation}
Naturally, the final radius, $r_f=r_p$, is fixed by the radial distance of the apex anchor from the Earth, $r_p$.
Note that variants of Equation~\eqref{eqn:vf_1} have appeared previously in~\cite{knudsen_thesis,mcinnes_2006,pearson_1997}. However, our derivation is more aligned with the one typically used to obtain Tsiolkovskii's rocket equation.

\subsection{Special case: Maximize Velocity by Releasing at GEO}\label{subsec:vf_max}
Equation~\eqref{eqn:vf_1} gives an expression for the radial velocity as a function of start radius, $r_0$, and the radius of the apex anchor, $r_f=r_p$. In this subsection, we determine the maximum possible radial velocity as a function of $r_p$. Specifically, $v(r_p,r_0)$ is maximized if $r_0=r_g$ is the geosynchronous radius or GEO - the balance point between centripetal acceleration and gravity. At $r_0=r_g$, we have
\[
-\frac{\mu}{r_0^2}+\omega_e^2 r_0=0
\]
or
\[
r_0=\sqrt[3]{\frac{\mu}{\omega_e^2}}=r_g.
\]
If we substitute this expression for $r_0$ into Equation~\ref{eqn:vf_1}, we find that we may simplify part of that expression as
\begin{align*}
-\frac{2\mu}{r_0}- \omega_e^2 r_0^2&=-\frac{1}{r_0}\left(2\mu + \omega_e^2 r_0^3\right)=-\frac{1}{r_0}\left(2\mu + \omega_e^2 \frac{\mu}{\omega_e^2}\right)=-\frac{1}{r_0}\left(2\mu + \mu \right)\\
&=-\frac{1}{r_0} 3\mu=-3\sqrt[3]{\frac{\omega_e^2}{\mu}} \mu=-3\sqrt[3]{\frac{\omega_e^2\mu^3}{\mu}} =-3\sqrt[3]{\omega_e^2\mu^2}\\
&=-3(\omega_e\mu)^{2/3}.
\end{align*}

\begin{figure}
  \centering
  \includegraphics[width=.5\textwidth]{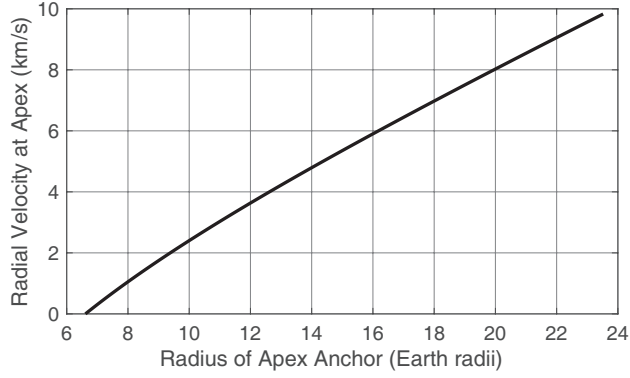}
  \caption{Radial velocity ($v_r$) at apex anchor for a sliding spacecraft started at GEO ($r_0=r_{GEO}$) as a function of radial distance to apex anchor, $r_p$ (Subtract 1 ER for space elevator length) with $v_0=0$.}\label{fig:vr}
\end{figure}

Now, by applying this simplification to Equation~\eqref{eqn:vf_1}, we obtain the Space Elevator Velocity Equation:
\begin{equation}\label{eqn:SE_rocket}
v_{r,\max}=\sqrt{v_0^2 + \frac{2\mu}{r_p}+\omega_e^2 r_p^2 -3(\omega_e\mu)^{2/3}}.
\end{equation}
Equation~\eqref{eqn:SE_rocket} gives the maximum radial velocity as a function of radial distance to the apex anchor, $r_p$ (which depends on elevator length). This radial velocity for $v_0=0$ is illustrated in Figure~\ref{fig:vr}. From this plot, we find that for $r_p=100,000km$, we have $v_r=5.705km/s$ and for $r_p=150,000km$, we have $v_r=9.7978km/s$. Note that these numbers do not include the tangential velocity of the apex anchor itself, which is $v_t=\omega_e r_p$

\subsection{Obtaining a desired terminal radial velocity}
In Subsection~\ref{subsec:vf_max}, we determined the maximum radial velocity as a function of radial distance to the apex anchor, $r_f=r_p$. In this subsection, we show that for a fixed $r_f=r_p$, the radial velocity, $v(r_p,r_0)$, is freely assignable up to this maximum velocity by appropriate choice of $r_0$. Furthermore, we obtain an analytic expression for the start radius, $r_d$ as a function of the desired radial release velocity, $v_d$. This ability to assign radial velocity magnitude is critical for all tiers of space elevator, as it adds a degree of freedom which can be used for targeting.

Specifically, for a given desired radial release velocity, $v_d$, by examining Equation~\eqref{eqn:vf_1}, we require the corresponding $r_d$ to satisfy
\[
v_d^2=\frac{2\mu}{r_f}+\omega_e^2 r_f^2-\frac{2\mu}{r_d}-\omega_e^2 r_d^2
\]
or
\begin{equation}\label{eqn:rd_eqn}
r_d^3+\underbrace{\left(\frac{v_d^2}{\omega_e^2}-\frac{2\mu}{\omega_e^2 r_f}-r_f^2\right)}_{p}r_d+\underbrace{\frac{2\mu}{\omega_e^2}}_{q}=0.
\end{equation}
This form of equation is a referred to as a depressed cubic and has an analytic solution given by Cardano's formula. Specifically, we have that for a given desired radial velocity, $v_d$, the associated start radius, $r_d$ is given by
\begin{equation}\label{eqn:rd_noramp}
r_d=\sqrt[3]{-\frac q2+\sqrt{\frac{q^2}4+\frac{p^3}{27}}} +\sqrt[3]{-\frac q2-\sqrt{\frac{q^2}4+\frac{p^3}{27}}},
\end{equation}
where $p$ and $q$ are as defined in Equation~\eqref{eqn:rd_eqn} and which is real-valued if $v_d\le v_{r,\max}$. We conclude that the radial velocity magnitude is freely assignable up to the maximum as determined by Equation~\eqref{eqn:SE_rocket}.

\section{Velocity Vector at Exit from Earth's SOI in GCRF Coordinates}\label{sec:velocity}
Now that we know the magnitude of the radial velocity of a sliding spacecraft at arrival at the apex anchor, we can perform some basic vector analysis of what happens at departure from the apex anchor and as the spacecraft leaves the SOI of the Earth. Complicating the analysis is the fact that one of three things can happen at the apex anchor, depending on the elevator design tier. In the first tier, nothing happens - there is no ramp and the velocity vector is unchanged. In the second tier, the apex ramp redirects the radial velocity in the direction of the tangential velocity - which is significantly more efficient, but does not allow for redirection into the ecliptic plane. In the third tier, the apex ramp is rotated by some amount, $\theta_r$, which can be used to track the ecliptic plane. In all cases, as the spacecraft exits the SOI of the Earth, the magnitude of the velocity decreases and the velocity vector bends slightly back towards the Earth by an amount we refer to as the turning angle, $\theta_{TA}$.

The goal, then, is to find the velocity vector in GCRF coordinates at exit from Earth's SOI in each case. In order to avoid repetition and keep this section brief, we will carefully describe the coordinate rotations from SEI to GCRF only for the first tier. Furthermore, we will leave our derivation of turning angle, $\theta_{TA}$ to later sections. Finally, recall that $R_1(\theta)$, $R_2(\theta)$, and $R_3(\theta)$ are the rotation matrices in Euclidean space, where positive $\theta$ is defined by the right hand rule.

\subsection{Velocity Vectors at Departure from the Apex Anchor}
Before we begin, we recall the definition of the Space Elevator Inertial (SEI) reference frame, as this coordinate system is not commonly used in existing literature. As depicted in Figures~\ref{fig:SEI} and~\ref{fig:SEI_polar}, in the SEI frame, the origin is at the center of the Earth, the $\hat x_{SEI}$ unit vector is defined as the outward radial of the elevator at release, $\hat z_{SEI}$ points towards the north pole, and $\hat y_{SEI}=\hat z \times \hat x$ points in the direction of the tangential velocity vector of the apex anchor itself. Given these definitions, we first define the velocity vector of the apex anchor itself, which is sometimes referred to as the tangential velocity,
\[
[\vec v_t]_{SEI}=\bmat{0\\ v_t\\0}=\bmat{0\\ \omega_e r_p\\0},
\]
where recall $\omega_e$ is the rotation rate of the Earth and $r_p$ is the radius of the apex anchor. Next, we have the initial radial velocity vector of the spacecraft at approach to the apex anchor:
\[
[\vec v_r]_{SEI}=\bmat{v_r\\ 0\\0}.
\]
Thus for a Tier 1 elevator with no apex ramp, the velocity vector at departure from the apex is
\[
[\vec v_{0-,T1}]_{SEI}=\bmat{v_r\\ \omega_e r_p \\0}.
\]
For a Tier 2 elevator, an apex ramp is present, with rotation angle, $\theta_c$. In this case, the velocity vector at departure from apex becomes
\[
[\vec v_{0-,T2}]_{SEI}=\bmat{0\\ \omega_e r_p\\0}+R_3(\theta_c)\bmat{v_r\\ 0\\0}.
\]
For a Tier 3 elevator, the apex ramp is rotated about the $\hat x_{SEI}$ axis. In this case, the velocity vector at departure from apex becomes
\[
[\vec v_{0-,T3}]_{SEI}=\bmat{0\\ \omega_e r_p\\0}+R_1(\theta_r)R_3(\theta_c)\bmat{v_r\\ 0\\0}.
\]

\subsection{Tier 1 Elevator: Velocity Vector at Exit from Earth's SOI in GCRF}
For Tier 1 and 2 elevators, the orbital plane after departure coincides with the equatorial plane, simplifying the analysis. Specifically, if $\theta_{TA}$ is the bend in the departure hyperbola from release to exit from the Earth's SOI, then for a Tier 1 elevator, the excess velocity vector at exit from Earth's SOI in SEI coordinates is
\[
[\vec v_{0+}]_{SEI}=\eta R_3(\theta_{TA})\bmat{v_r\\\omega_e r_p\\0}
\]
where
\[
\eta=\sqrt{\frac{v_{exc}}{v_r^2+\omega_e^2 r_p^2}}
\]
is the factor by which the magnitude of the velocity decreases along the departure hyperbola. The expressions for $v_{exc}$ and $\theta_{TA}$ will be derived for a Tier 1 elevator in Section~\ref{sec:thc=0}.

To convert from SEI to Earth Centered Earth Fixed (ECEF) coordinates, we assume the base anchor is located on the equator and that the longitude of this facility is $\lambda_{SE}$ - which defines the orientation of the $\hat x_{SEI}$ unit vector. The $\hat z_{ECEF}$ unit vector aligns with the $\hat z_{SEI}$ unit vector. Since the $\hat x_{ECEF}$ unit vector is defined by the Greenwich meridian, the ECEF frame is obtained from the SEI frame by a negative rotation about the $\hat z_{SEI}$ axis by amount $\lambda_{SE}$, so that
\[
[\vec v_{0+}]_{ECEF}=\eta R_3(\lambda_{SE})R_3(\theta_{TA})\bmat{v_r\\\omega_e r_p\\0}.
\]
In the ECI coordinate system, the $\hat z_{ECI}$ unit vector aligns with that of the SEI and ECEF frames. The $\hat x_{ECI}$ unit vector, however, points to the FPOA and is obtained from the $\hat x_{ECEF}$ unit vector by a negative rotation by the angle defined by the Earth Rotation Angle (ERA) or Greenwich Mean Time (GMT), denoted $\theta_{ERA}(t)$. Thus
\[
[\vec v]_{ECI}=R_3(\theta_{ERA}(t))[\vec v]_{ECEF}.
\]
For convenience, we combine $\lambda_{SE}$ and $\theta_{ERA}(t)$ into a single rotation angle $\theta_{LST}(t)=\lambda_{SE}+\theta_{ERA}(t)$, which represents Local Sidereal Time (LST) at the base of the space elevator. Thus we have
\[
[\vec v_{0+}]_{ECI}=\eta R_3(\theta_{LST})R_3(\theta_{TA})\bmat{v_r\\\omega_e r_p\\0}.
\]
Finally, in the Geocentric Celestial Reference Frame (GCRF), the $\hat x_{\epsilon}$ unit vector points to the FPOA and is aligned with $\hat x_{ECI}$. However, the $\hat z_{\epsilon}$ unit vector points to the North Celestial Pole (ecliptic-normal) and is thus obtained from $\hat z_{ECI}$ by a positive rotation about the $\hat x_{ECI}$ axis by the inclination (obliquity) to the ecliptic, $\epsilon$ - See Figure~\ref{fig:SEI}. Thus the velocity vector at departure from the Earth SOI in GCRF coordinates for a Tier 1 elevator (with no apex ramp) is
\begin{equation}\label{eqn:vexc_GCRF_noramp}
[\vec v_{0+,T1}]_{GCRF}=\eta R_1(-\epsilon)R_3(\theta_{LST}+\theta_{TA})\bmat{v_r\\\omega_e r_p\\0}.
\end{equation}
\subsection{Tier 2 Elevator: Velocity Vector at Exit from Earth's SOI in GCRF}
The analysis of an elevator with a ramp but no rotation is similar to the case without a ramp, in that the orbital plane and the equatorial plane coincide. Specifically, we have
\begin{equation}\label{eqn:vexc_GCRF_ramp_norot}
[\vec v_{0+,T2}]_{GCRF}=\eta R_1(-\epsilon)R_3(\theta_{LST}+\theta_{TA})\left(\bmat{0\\ \omega_e r_p\\0}+R_3(\theta_c)\bmat{v_r\\ 0\\0}\right).
\end{equation}
The expressions for $\eta$ and $\theta_{TA}$ will be derived for the Tier 2 elevator in Section~\ref{sec:escape}.
\subsection{Tier 3 Elevator: Velocity Vector at Exit from Earth's SOI in GCRF}
Unfortunately or fortunately, launch from a Tier 3 elevator is more complicated because the rotation of the apex ramp changes the orientation of the orbital plane of the departing spacecraft. To simplify the analysis, we assume $\theta_c=90^\circ$. Other than survivability, there is no obvious benefit to a smaller angle of redirection. Moreover, this assumption allows us to conclude that the periapse of the departure hyperbola occurs at release from the space elevator. This allows us to define the orientation of the perifocal (PQW) frame, where the $\hat x_{PQW}=\hat p$ unit vector aligns with $\hat x_{SEI}$, while $\hat z_{PQW}=\hat w$ aligns with $[\vec v_{0-}]_{SEI} \times \hat x_{SEI}$. Hence, we are able to begin our analysis in the perifocal frame with
\[
[\vec v_{0-}]_{PQW}=\bmat{0\\v_p\\0}_{PQW},
\]
where (since $\theta_c=90^\circ$),
\begin{align*}
v_p&=\norm{[\vec v_{0-,T3}]_{SEI}}\\
&=\sqrt{\omega_e^2r_p^2+2\omega_e r_p v_r \cos(\theta_r)+v_r^2}.
\end{align*}
In the PQW frame, as the spacecraft moves on its departure hyperbola, the velocity vector rotates positively by amount $\theta_{TA}$ about the $\hat z_{PQW}$ axis. In addition, the magnitude of the velocity vector will decrease along the departure hyperbola from $v_{p}$ to $v_{exc}$, where
\[
v_{exc}=\sqrt{-\frac{\mu}{a}}=\sqrt{v_p^2-\frac{2\mu}{r_p}}.
\]
Therefore, the velocity vector at departure from the Earth's SOI in the PQW frame is
\[
[\vec v_{0+}]_{PQW}=R_{3}(\theta_{TA})\bmat{0\\v_{exc}\\0}.
\]
Now all that remains is to convert from PQW coordinates to SEI and hence to GCRF. However, this is complicated because we must use $\theta_r$ to determine the orientation of the orbital plane relative to the equatorial plane.
Specifically, since $\theta_{c}=90^\circ$, $\hat x_{PQW}=\hat x_{SEI}$, and so the SEI frame is obtained from the PQW frame by a negative rotation about the $\hat x$ axis by angle $\theta_v$ where
\begin{equation}\label{eqn:theta_v}
\theta_v=\tan^{-1}\frac{[\vec v_{0-}]_{SEI}\cdot \hat z_{SEI}}{[\vec v_{0-}]_{SEI}\cdot \hat y_{SEI}}=\tan^{-1}\frac{v_r \sin(\theta_r)}{\omega_e r_p +v_r \cos(\theta_r)}.
\end{equation}
The velocity vector in the SEI frame at departure from the Earth SOI is thus
\begin{equation}\label{eqn:vexc_SEI}
[v_{0+}]_{SEI}=R_{1}(\theta_v)R_{3}(\theta_{TA})\bmat{0\\v_{exc}\\0}.
\end{equation}
Hence in GCRF coordinates, we have
\begin{equation}\label{eqn:vexc_GCRF}
[\vec v_{0+,T3}]_{GCRF}=R_1(-\epsilon)R_3(\theta_{LST}) R_{1}(\theta_v)R_{3}(\theta_{TA})\bmat{0\\v_{exc}\\0}.
\end{equation}
Note that when $\theta_r=0$, we have $\theta_v=0$ and hence we recover the Tier 2 velocity vector. In the following Sections, we will use Equations~\eqref{eqn:vexc_GCRF_noramp},~\eqref{eqn:vexc_GCRF_ramp_norot}, and~\eqref{eqn:vexc_GCRF} to determine the limits of the ability of the space elevator to enable propellant-free travel to the distant planets.

\section{Restriction to the Ecliptic Plane}\label{sec:ecliptic_motivation}
At this point it is necessary to confront an issue which will arise repeatedly throughout the paper - transfer to the ecliptic plane.

All planets in the solar system share a single orbital plane - Laplace's invariable plane, with each planetary orbit at some slight inclination to this plane, an inclination which is typically less than $2^\circ$. The Earth's orbital plane, for example, is inclined by $1.57^\circ$, and is referred to as the ecliptic plane (named after the eclipses, all of which occur in it). For our purposes, however, we will assume that the ecliptic plane, the invariable plane, and the orbital planes of the distant planets all roughly coincide and use the term ecliptic to refer to this somewhat fictitious plane.

In this framework, then, any spacecraft launched from the Earth to a distant planet must necessarily have initial and final position vectors located in the ecliptic plane. This implies that any transfer orbit between the Earth and a distant planet must necessarily likewise lie entirely in the ecliptic plane - with the very rare exception of the case when initial and final positions are co-linear. Therefore, if a spacecraft launched from the Earth is required to arrive at a distant planet, a necessary condition for a successful intercept is that the transfer orbit lie in the ecliptic plane.

The necessity of ecliptic-bound outgoing transfer orbits is unfortunate for space elevator advocates because the motion of the space elevator lies entirely within the equatorial plane, which is inclined with respect to the ecliptic plane by a substantial $23.4^\circ$ (also known as the obliquity to the ecliptic - See Figure~\ref{fig:SEI}). Note that this unfortunate reality is due to centripetal acceleration (which is perpendicular to the rotation axis of the Earth) and not geography and so would still hold even if the elevator were built at, e.g. northern latitudes. Thus a space elevator built at $23.4^\circ$ north latitude would appear to an observer on the ground to be leaning off to the south at an angle of $23.4^\circ$.

\paragraph{The Orbital Plane of Departure in Earth's SOI Can Not Be the Ecliptic} Given the geometry of the space elevator, the first thing to note is that it is not possible to restrict ourselves to departure orbits which lie in the ecliptic plane while still in the Earth's SOI. To explain why this is impossible, let us consider what happens when a spacecraft is launched from the apex of the space elevator. The orbital plane of the outgoing spacecraft with respect to the Earth is determined by the angular momentum vector,
\[\vec h_{GCRF}=\vec r_{GCRF} \times \vec v_{GCRF}=\bmat{h_x\\h_y\\h_z}.
\]
The ecliptic plane, meanwhile is defined by the vector $[\vec h_{\epsilon}]_{GCRF}=\bmat{0&0&1}^T$. So, at this point, in order for the outgoing orbit to lie in the ecliptic plane, we would require $[\vec h_{\epsilon}]_{GCRF}\times \vec h_{GCRF} =0$ or $h_x=h_y=0$. This constraint implies that both the position and velocity vectors at release are orthogonal to $\vec h_{z}$ - i.e. $v_z=r_z=0$.  We regretfully conclude, therefore, that this constraint is impossible to satisfy using Tier 1 or Tier 2 space elevators. This conclusion is based on the fact that without rotation of the ramp, both the position and velocity vectors lie in the equatorial plane and are not co-linear. Note that even with a rotation of the ramp, for an ecliptic orbital departure plane, launch would necessarily have to occur at the ascending or descending node, a window which only occurs twice per day.

\paragraph{The Orbital Plane of Departure in Earth's SOI Doesn't Matter} Fortunately, the orbital plane of the departure hyperbola while in Earth's SOI does not actually determine the orbital plane after departure from the SOI - only the velocity vector at exit from Earth's SOI matters. This is because the orbital plane after departure from the Earth is determined in Barycentric coordinates as $\vec h_{BCRS}=\vec r_{BCRS} \times \vec v_{BCRS}$. This is significant because the origin in the BCRS coordinate system is the Sun and hence
\[
\vec r_{BCRS}=\vec r_{GCRF}+\vec r_{s\rightarrow e}
\]
where $\vec r_{s\rightarrow e}$ is the Sun-Earth vector in BCRS with magnitude equal to the distance from the Sun to the Earth, which is approximately three orders of magnitude larger than $\norm{\vec r_{GCRF}}$ at exit from the SOI. Thus, when computing $\vec h_{BCRS}$, we may neglect $\vec r_{GCRF}$, implying that the position vector, $\vec r_{BCRS}\cong\vec r_{s\rightarrow e}$ already lies in the ecliptic plane. Under this assumption, a necessary and sufficient condition for $\vec h_{BCRS} \times \vec h_{\epsilon}=0$ is for the excess velocity vector in GCRF coordinates to satisfy $[\vec v_{exc}]_{GCRF}\cdot \vec h_{\epsilon}=0$. Specifically, the $\hat z$ component of $[\vec v_{exc}]_{GCRF}$ must be zero.

This difference between two constraints on the outgoing orbital plane and a single constraint on the terminal velocity vector is important, since none of our elevator designs allow for more than 3 degrees of freedom - $v_r$, $\theta_{LST}$ and $\theta_r$.  Specifically, using the relaxed constraint $[\vec v_{exc}]_{GCRF}\cdot \hat z=0$, as shown in Section~\ref{sec:ecliptic}, for a Tier 3 space elevator of apex radius greater than $60,579km$, an appropriate choice of $\theta_r$ always exists for launch into the ecliptic, regardless of launch time. Furthermore, as discussed in Sections~\ref{sec:thc=0} and~\ref{sec:escape}, for a Tier 1 or 2 space elevator, transition to the ecliptic is still possible by appropriate choice of release time, $\theta_{LST}$. Specifically, for a Tier 2 elevator, a free release at time $\theta_{LST}=90^\circ-\theta_{TA}$ or $\theta_{LST}=270^\circ-\theta_{TA}$ results in transfer to the ecliptic, where recall $\theta_{TA}$ is the turning angle of the departure hyperbola.

Having defined the problem of transfer to the ecliptic plane, we next study free release transfers to the distant planets for a Tier 1 Space Elevator.

\section{Tier 1 Space Elevator: Free Release Transfer without Apex Ramp}\label{sec:thc=0}
We begin our analysis of the orbital mechanics of space elevators with the Tier 1 class. These elevators are the simplest to construct in that they do not require apex ramps and have been the basic design used in all previous studies.
The disadvantage of the Tier 1 class over Tier 2, of course, is the inefficient use of radial velocity - resulting in the necessity to extend the length of the elevator to achieve the same transfer orbits. Furthermore, for both Tier 1 and 2 space elevators, we have only two degrees of freedom, $v_r$ and $\theta_{LST}$ - we do not yet consider the use of an apex ramp rotation, $\theta_r$, as in Sections~\ref{sec:ecliptic}-\ref{sec:lambert}. However, as we will see, Tier 1 and 2 space elevators are still capable of launching free release transfers to the outer planets, albeit much less frequently than with an apex rotation. Note that we do not seriously consider elevators without a sliding surface (Tier 0), as free transfer opportunities using such elevators are practically non-existent.

To begin this section, we determine the minimum length of a Tier 1 space elevator for which a spacecraft, released at the apex anchor, will escape the SOI of the Earth - a number which is then compared to Tier 2 elevators in Section~\ref{sec:escape}. Furthermore, for elevators which exceed this minimum length, we also determine the associated excess velocity. Next, in Subsection~\ref{subsec:T1_releasetime}, we calculate the release time, $\theta_{LST}$, necessary for free transfer to the ecliptic plane. In Subsections~\ref{subsec:T1_vexc_desired} and~\ref{subsec:T1_hohmann}, we then compute the minimum elevator length for free release Hohmann transfer to the distant planets. Finally, in Subsection~\ref{subsec:T1_nonhohmann} we show that by appropriate choice of start radius (and hence $v_r$), we can dramatically increase the frequency of free release opportunities by considering non-Hohmann transfers.
%

\subsection{Radial Velocity and Excess Velocity}\label{subsec:T1_vr_vexc}
To begin, we examine the total energy of a spacecraft after release and deduce the minimum elevator length for escape from Earth's SOI and the resulting maximum excess velocity magnitude for elevators which exceed this minimum length.
First, we set $\theta_c=\theta_r=0$, $v_0=0$, $r_0=r_g$ and we determine the smallest $r_p$ such that the total energy in the SOI of the Earth is positive. This total energy, $V$, is a combination of gravitational potential and kinetic energy:
\[
V=-\frac{\mu_e}{r_p}+\frac{\norm{\vec v}^2}{2},
\]
where $\vec v$ is the velocity at release in any inertial reference frame and $r_p$ is the apex radius. Using the Space Elevator Inertial (SEI) coordinate system we have
\[
\vec v=\bmat{v_r\\v_t\\0}_{SEI}.
\]
Thus the magnitude of the velocity vector after release is
\begin{align*}
\norm{v}^2&=v_t^2+v_r^2=\omega_e^2 r_p^2+v_r^2=\omega_e^2 r_p^2+\frac{2\mu}{r_p}+\omega_e^2 r_p^2 -3(\omega_e\mu)^{2/3}\\
&=2\omega_e^2 r_p^2+\frac{2\mu}{r_p} -3(\omega_e\mu)^{2/3}.
\end{align*}
Now solving for the potential energy, we have
\begin{align*}
2V&=-\frac{2\mu}{r_p}+\norm{v}^2\\
&=-\frac{2\mu}{r_p}+\frac{2\mu}{r_p}+2\omega_e^2 r_p^2 -3(\omega_e\mu)^{2/3}\\
&=2\omega_e^2 r_p^2 -3(\omega_e\mu)^{2/3}.
\end{align*}
Therefore, the minimal $r_p$ such that $V\ge 0$ can be found in a closed-form expression as
\begin{align*}
 r_{f,\min} &=\sqrt{\frac{3}{2}}\mu^{2/6}\omega_e^{-4/6}\\
&=\sqrt{\frac{3}{2}}\sqrt[3]{\frac{\mu}{\omega_e^2}}=51,640.
\end{align*}

Having obtained the shortest space elevator for which we can escape the SOI of the Earth, we next determine the associated excess velocity magnitude for a spacecraft released from a Tier 1 elevator which exceeds this minimum length. Recall that the excess velocity is the velocity magnitude after the craft has left the gravity well of the planet (Earth, in this case) and is given by $v_{exc}=\sqrt{2V}$. Therefore, for a radial release (Tier 1 design), we have the following expression:
\begin{equation}
v_{exc,\max}=\sqrt{2\omega_e^2 r_p^2 -3(\omega_e\mu)^{2/3}}.\label{eqn:T1_vexc_rad_max}
\end{equation}
This excess velocity magnitude is illustrated in Figure~\ref{fig:vexc}, which includes a comparison with the non-sliding case (Tier 0) and with a Tier 2 or 3 elevator.
In the case where an excess velocity smaller than the maximum is desired, we may use the following formula which gives $v_{exc}$ as a function of $v_r$.
\begin{equation}\label{eqn:T1_vexc_vr}
v_{exc}=\sqrt{\norm{v}^2-\frac{2\mu}{r_p}}=\sqrt{\omega_e^2 r_p^2+v_r^2-\frac{2\mu}{r_p}}
\end{equation}
Recall that any $v_r\le v_{r,\max}$ may be selected using start radius, $r_0$, as given by Equation~\eqref{eqn:rd_noramp}.
\paragraph{Comparison with a Non-Sliding Spacecraft (Tier 0 elevator)}
The failure to use centripetal acceleration to produce a radial velocity component significantly reduces the capabilities of the space elevator for deep space exploration. Specifically, if we assume $v_r=0$ (or $r_0=r_p$), then the magnitude of the velocity vector decreases to
\[
\norm{\vec v}^2=v_t^2=\omega_e^2 r_p^2.
\]
If we substitute this value into the expression for total energy, we have
\[
2V=-\frac{2\mu}{r_p}+\omega_e^2 r_p^2.
\]
Thus the shortest space elevator to escape the Earth's SOI has apex radius
\begin{align*}
&r_{f,\min}=\sqrt[3]{2}\sqrt[3]{\frac{\mu}{\omega_e^2}}=53,123.
\end{align*}
So in this case, we require an additional 1,483 km of space elevator. Furthermore, for elevators which exceed this length, we have the following reduced expression for excess velocity:
\begin{equation}\label{eqn:vexc_norad}
v_{exc,\max}=\sqrt{\omega_e^2 r_p^2-\frac{2\mu}{r_p}}.
\end{equation}
This excess velocity magnitude is also illustrated in Figure~\ref{fig:vexc} and shows an approximately 50\% reduction in excess velocity magnitude when compared to a tangential release (examined in Section~\ref{sec:escape}). More significantly, however, the magnitude of this excess velocity is permanently fixed, which, as we will see, practically eliminates our ability to obtain free release transfers to the outer planets.

\subsection{Computing Release Time for Free Transfer to Ecliptic}\label{subsec:T1_releasetime}
In Section~\ref{sec:velocity}, we determined the velocity vector in GCRF coordinates at exit from Earth's SOI. As shown in Section~\ref{sec:ecliptic_motivation}, for transfer to distant planets we must ensure the velocity vector at exit from Earth's SOI in GCRF coordinates lies in the ecliptic plane - Specifically, $[\vec v_{0+,T1}]_{GCRF}\cdot \hat z=0$. From Section~\ref{sec:velocity} we have
\begin{align*}
[\vec v_{0+,T1}]_{GCRF}&=\eta R_1(-\epsilon)R_3(\theta_{LST}+\theta_{TA})\bmat{v_r\\\omega_e r\\0}\\
&=\eta R_1(-\epsilon)\bmat{v_r \cos (\theta_{LST}+\theta_{TA})-\omega_e^2 r_p^2 \sin (\theta_{LST}+\theta_{TA})\\v_r \sin (\theta_{LST}+\theta_{TA})+\omega_e^2 r_p^2 \cos (\theta_{LST}+\theta_{TA})\\0}.
\end{align*}
where $\eta=\frac{v_{exc}}{\sqrt{v_r^2+\omega_e^2 r_p^2}}$ and $v_r$ is given by Equation~\eqref{eqn:rd_noramp}. This implies that a necessary and sufficient condition for $[\vec v_{0+,T1}]_{GCRF}\cdot \hat z=0$  is
\[
v_r \sin (\theta_{LST}+\theta_{TA})+\omega_e^2 r_p^2 \cos (\theta_{LST}+\theta_{TA})=0,
\]
which has two solutions given by
\begin{equation}\label{eqn:thLST_noramp}
\theta_{LST}=\tan^{-1}\left(-\frac{\omega_e r_p}{v_r}\right)-\theta_{TA}+180^\circ\quad \text{and}\quad \theta_{LST}=\tan^{-1}\left(-\frac{\omega_e r_p}{v_r}\right)-\theta_{TA}+360^\circ.
\end{equation}
The corresponding velocity vectors in GCRF at exit from the Earth's SOI are thus
\begin{equation}
[\vec v_{0+,T1}]_{GCRF}=\bmat{\pm \sqrt{v_r^2+\omega_e^2r_p^2-2 \frac{\mu}{r_p} }\\0\\0}=\bmat{\pm v_{exc}\\0\\0}
\end{equation}
where again $v_r$ is given by Equation~\eqref{eqn:rd_noramp} and $v_{exc}$ is given by Equation~\eqref{eqn:T1_vexc_vr}.

Thus we conclude that there are exactly two release times per day for which the velocity vector at exit from the Earth's SOI lies in the ecliptic plane. Furthermore, the corresponding exit velocity vectors in GCRF coordinates are \textit{permanently fixed}, in that their direction does not depend on launch date, start radius ($r_0$), or apex radius ($r_p$). Note, however, that $r_0$ and $r_p$ do affect the \textit{magnitude} of the exit velocity vector - a degree of freedom we will exploit in Subsection~\ref{subsec:T1_nonhohmann}.

\paragraph{Computing Turning Angle, $\theta_{TA}$} Finally, we note that to calculate the free release times, we additionally need to determine $\theta_{TA}$ - the turning angle of the departure hyperbola after release. For this calculation, however, we need to first determine the corresponding orbital element, eccentricity, $e$. For this, we note that the velocity and position vectors at release are
\[
[\vec v_{0+}]_{SEI}=\bmat{v_r\\\omega_e r\\0},\qquad [\vec r_{0+}]_{SEI}=\bmat{r_p\\0\\0}.
\]
The angular momentum vector is thus
\[
\vec h =[\vec r_{0+}]_{SEI} \times [\vec v_{0+}]_{SEI}=\bmat{0\\0\\\omega_e r_p^2}.
\]
The eccentricity vector is then
\begin{align*}
\vec e&=\frac{1}{\mu}\vec v \times \vec h-\frac{\vec r}{\norm{\vec r}}\\
&=\frac{1}{\mu}\bmat{v_r\\\omega_e r\\0} \times \bmat{0\\0\\\omega_e r_p^2}-\bmat{1\\0\\0}=\bmat{\frac{1}{\mu}\omega_e^2r_p^3-1\\-\frac{1}{\mu}v_r\omega_e r_p^2\\0}.
\end{align*}
The eccentricity is the magnitude of the eccentricity vector which can be found as
\[
e=\sqrt{\left(\frac{1}{\mu}\omega_e^2r_p^3-1\right)^2+ \left(\frac{1}{\mu}v_r\omega_e r_p^2\right)^2}
\]
with the resulting turning angle given by
\[
\theta_{TA,tot}=\sin^{-1}\left(\frac{1}{e}\right).
\]
Note, however, that $\theta_{TA,tot}$ is the turning angle from periapse and for a Tier 1 elevator, periapse does not occur at release, but rather occurs earlier on some fictional orbit prior to release. Hence, we must also calculate the argument of periapse at release from
\[
\nu_{release}=\cos^{-1}\left(\frac{\vec r \cdot \vec e}{r_p e}\right)=\cos^{-1}\left(\frac{r_p\left(\frac{1}{\mu}\omega_e^2r_p^3-1\right)}{r_p e}\right).
\]
Thus the actual turning angle from release to exit from Earth's SOI is given by
\[
\theta_{TA}=\theta_{TA,tot}-\nu_{release}=\sin^{-1}\left(\frac{1}{e}\right)-\cos^{-1}\left(\frac{r_p\left(\frac{1}{\mu}\omega_e^2r_p^3-1\right)}{r_p e}\right).
\]

\subsection{Space Elevator Length as a Function of Desired Excess Velocity}\label{subsec:T1_vexc_desired}
Having shown that transfer to the ecliptic is possible by choice of launch time, $\theta_{LST}$, we now examine the ability of the space elevator to provide Hohmann transfers to the outer planets assuming perfect alignment. For this calculation, however, we first need to use Equation~\eqref{eqn:T1_vexc_rad_max} to find an analytic expression for the apex anchor radius, $r_p$ which achieves a given desired velocity, $v_d=v_{exc}$, at exit from the Earth's SOI. Note that this analysis is different from that used to obtain Equation~\eqref{eqn:rd_noramp}, wherein we found the \textit{start radius} to obtain a desired radial velocity, $v_r$ at apex. Specifically, we now assume that start is at GEO ($r_0=r_g$) and determine the apex radius such that $v_{exc}=v_d$ for a given desired excess velocity, $v_d$, based on the formula for excess velocity in Eqn.~\eqref{eqn:T1_vexc_rad_max} in Subsection~\ref{subsec:T1_vr_vexc}. Specifically, we require
\[
v_d^2=2\omega_e^2 r_p^2 -3(\omega_e\mu)^{2/3}
\]
or
\begin{equation}\label{eqn:vdes_noramp}
r_p =\sqrt{\frac{1}{2\omega_e^2}\left(v_d^2+3(\omega_e\mu)^{2/3}\right)}.
\end{equation}
This solution yields the apex radius, $r_p$, for which the excess velocity at exit from Earth's SOI is $v_d$. The corresponding space elevator length is obtained by subtracting one Earth radius. In the following subsection, we use Equation~\eqref{eqn:vdes_noramp} to obtain the minimum Tier 1 elevator length for free transfer to the outer planets using a Hohmann transfer under ideal planetary alignment.

\subsection{Minimum Elevator Length for Hohmann Transfer to Distant Planets}\label{subsec:T1_hohmann}
In this subsection, we assume ideal planetary alignment and determine the minimum Tier 1 elevator length under which transfer to the distant planets is possible. Specifically, in this subsection, we first convert our velocity vector at exit from Earth's SOI to BCRS coordinates. Specifically, if $\lambda_e$ is the celestial longitude of the Earth as measured from the FPOA or vernal equinox, then the position and velocity vectors of the spacecraft at exit from Earth's SOI are
\[
[\vec r_{0+,T1}]_{BCRS}=R_3(\lambda_e)\bmat{d_e\\0\\0}\qquad \text{and}\qquad [\vec v_{0+,T1}]_{BCRS}=R_3(\lambda_e)\bmat{0\\v_e\\0}+\bmat{\pm v_{exc}\\0\\0}
\]
where $d_e$ is the distance of the Earth from the Sun and $v_e$ is the mean velocity of the Earth in its orbit. Thus for a Hohman transfer, we require $\lambda_e=90^\circ$ or $\lambda_e=270^\circ$ so that
\[
[\vec v_{0+,T1}]_{BCRS}=\bmat{\pm (v_{exc}+v_e)\\0\\0}.
\]
That is, we require the launch date to be either the winter or summer solstice. Furthermore, we require the target planet to have celestial longitude $\lambda_p=270^\circ$ at arrival for a winter launch or $\lambda_p=90^\circ$ at arrival for a summer launch. Obviously the precise planetary alignment required for such a Hohmann transfer is essentially unobtainable. However, we will see in Subsection~\ref{subsec:T1_nonhohmann} that this constraint can be relaxed somewhat if the elevator exceeds the minimum length for a Hohmann transfer.

For computing the minimum elevator length itself, we must first compute the excess velocity required to execute the Hohmann transfer to each given planet. Since $\Delta v$ calculations for the Hohmann transfer are well known, we do not include the details of these calculations. Instead, we simply apply Equation~\eqref{eqn:vdes_noramp} to the excess velocity required for Hohmann transfer to each planet minus the velocity of the Earth in the BCRS frame and tabulate the results in Table~\ref{tab:escape}, wherein the values are compared to those required for a Tier 2 elevator. For solar escape, the desired exit velocity in BCRS is the escape velocity from the Sun's SOI $\left(\sqrt{2\frac{\mu_s}{d_e}}\right)$, where $\mu_s$ is the gravitational parameter of the Sun and $d_e$ is the distance of the Earth from the Sun. 

\paragraph{Note on non-Ecliptic Transfers} As a final note before proceeding to non-Hohmann transfers, we should mention that Hohmann transfers are unique in that the initial and final position vectors are \textit{co-linear}. This means that technically, in this very special case, we may relax the constraint that the transfer orbit lie in the ecliptic plane. However, before getting too excited, Hohmann transfers do impose the additional constraint that the excess velocity vector in GCRF be orthogonal to the Earth-Sun axis. This additional constraint again uniquely determines the launch time and imposes very demanding planetary alignment constraints similar to those discussed earlier. Furthermore, these alignment constraints cannot be relaxed, as we do in the following Subsection. For these reasons, we do not pursue an analysis of non-ecliptic Hohmann transfers in this manuscript.


\subsection{Launch Windows for non-Hohmann Transfers}\label{subsec:T1_nonhohmann}
\begin{figure}[ht]
    \begin{subfigure}[t]{\textwidth}
  \centering
  \includegraphics[width=.75\textwidth]{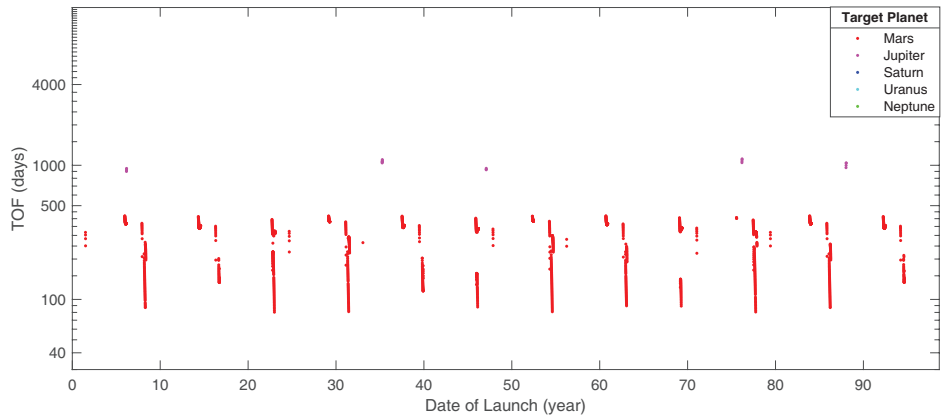}
  \caption{Times of Flight and launch dates for an apex radius of 100,000km.}\label{fig:TOFs_noramp_100000}
    \end{subfigure}
    \newline
    \begin{subfigure}[t]{\textwidth}
  \centering
  \includegraphics[width=.75\textwidth]{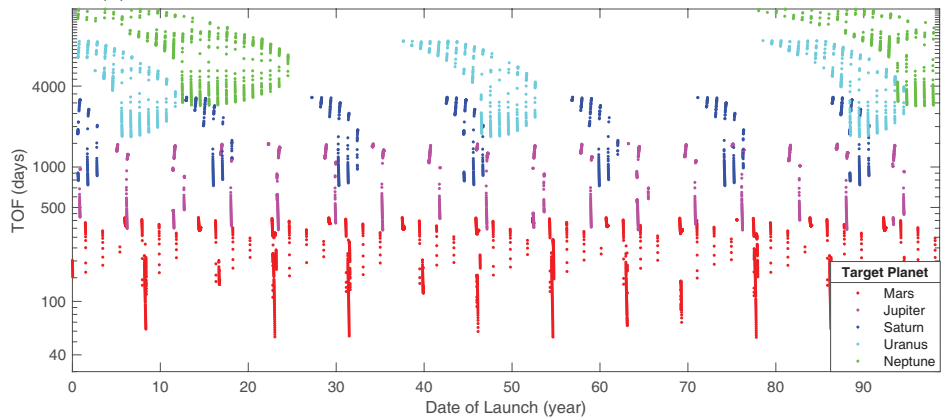}
  \caption{Times of Flight and launch dates for an apex radius of 150,000km.}\label{fig:TOFs_noramp_150000}
    \end{subfigure}
    \caption{Times of Flight and launch dates to for free release transfer to the outer planets for a Tier 1 space elevator over a 100 year interval. Each dot indicates a date on which a free release transfer is possible.}\label{fig:TOFs_noramp} \vspace{-5mm}
\end{figure}

In Subsection~\ref{subsec:T1_hohmann}, we showed that while Hohmann transfers are possible for Tier 1 elevators, the required planetary alignment renders them essentially unusable. This is because the restriction to Hohmann transfers uniquely determines the radial velocity and restriction to the ecliptic plane uniquely determines the launch time, $\theta_{LST}$, thus eliminating both our degrees of freedom. However, in this final subsection we note that if we relax the transfer orbit to be non-Hohmann, then we still have some flexibility in the magnitude of the radial component of our launch velocity. Specifically, recall that for $\theta_{LST}$ as given in Equation~\eqref{eqn:thLST_noramp}, and given $\lambda_E$, the position and velocity vectors in BCRS at departure from Earth's SOI are given by
\[
[\vec r_{0+,T1}]_{BCRS}=R_3(\lambda_e)\bmat{d_e\\0\\0}\qquad \text{and}\qquad [\vec v_{0+,T1}]_{BCRS}=R_3(\lambda_e)\bmat{0\\v_e\\0}+\bmat{\pm v_{exc}\\0\\0}
\]
where by selecting the start radius $r_0=r_d$ as in Equation~\eqref{eqn:rd_noramp}, we may vary the excess velocity magnitude, $v_{exc}$, in the interval
\[
v_{exc}=\sqrt{v_r^2+\omega_e^2r_p^2-\frac{2\mu}{r_p}}\in \left[\sqrt{\omega_e^2r_p^2-\frac{2\mu}{r_p}},\sqrt{2\omega_e^2r_p^2-3(\omega_e \mu)^{2/3}}\right].
\]
Thus for each $v_{exc}$ and $\lambda_e$, we have a departure orbit with associated orbital elements. Furthermore, for each departure orbit which reaches the radius of the target planet, we may calculate an associated date of arrival based on the polar equation (for elliptic orbits there will be two arrival dates). By using the mean motion of the target planet, we may then calculate the dates over a specified period of time which correspond to feasible transfers. These dates are listed in Figures~\ref{fig:TOFs_noramp_100000} and~\ref{fig:TOFs_noramp_150000} for apex radii of 100,000km and 150,000km respectively, along with associated Times of Flight (TOF) to each of the outer planets over a specified 100 year interval - each dot indicates a day on which transfer is possible along with the TOF for that transfer. While formulae for all of the orbital elements of the departure orbit and the corresponding arrival time may be obtained analytically (we do not need to solve Kepler's equation), these calculations are standard and are not particularly illuminating, and hence we omit them from this discussion. 

Thus referring to Figures~\ref{fig:TOFs_noramp_100000} and~\ref{fig:TOFs_noramp_150000}, we see that, as expected, there are no transfers to the outer planets beyond Jupiter for a Tier 1 space elevator of apex radius $r_p=100,000km$. Furthermore, even with the additional flexibility of non-Hohmann orbits, transfer to Jupiter only occurs approximately 5 times in 100 years. For a Tier 1 elevator of apex radius $r_p=150,000km$, meanwhile, more regular transfers are possible, although there are only 2 windows for Neptune and 3 for Uranus over this 100 year time period. Windows for Saturn and Juptier are more regular, with regular gaps of approximately 15 and 8 years, respectively. Gaps in transfer to Mars, meanwhile are only approximately 2-3 years. These gaps arise due to the fact that planetary alignment must occur near the winter or summer solstice - hence the vertical striations we observe in the launch windows. Even for Mars, availability never exceeds approximately once per year in our free transfer opportunities. These gaps, while reduced, are present even with Tier 2 elevators (See Figures~\ref{fig:TOFs_norotation_100000} and~\ref{fig:TOFs_norotation_150000}) and significantly reduce the benefits of these space elevator designs when considering the massive costs involved in construction. Thus in Sections~\ref{sec:ecliptic} and~\ref{sec:lambert} we consider the benefits of rotating the apex anchor to track the ecliptic. The main benefit of this modification is that it obviates the need for planetary alignment to coincide with the solstices.

Finally, we note that without a sliding spacecraft, the excess velocity cannot be adjusted using start radius, and hence the Tier 0 elevator is not considered in our analysis.
%

\section{Tier 2 Space Elevator: Free Release Transfer with Apex Ramp}\label{sec:escape}
In this section, we revisit the analysis of Section~\ref{sec:thc=0} while accounting for the increase in excess velocity and change of direction achieved by using an apex ramp, but without including the benefits associated with rotation of that ramp. Specifically, the Tier 2 elevator design includes an apex ramp with $\theta_c=90^\circ$, thereby ensuring the radial velocity aligns with the tangential velocity of the apex anchor itself. This assumption maximizes excess velocity and simplifies the calculation of turning angle of the departure hyperbola by ensuring that the periapse of the hyperbolic departure occurs at the apex anchor.

First, in Subsection~\ref{subsec:vexc_notheta}, we determine the excess velocity after departure from the apex ramp and show that this excess velocity significantly exceeds those values calculated for a Tier 1 design. Next, in Subsection~\ref{subsec:TA1}, we determine release time, $\theta_{LST}$, required for transition to the ecliptic plane using a simplified expression for the turning angle, $\theta_{TA}$ of the departure hyperbola. In Subsection~\ref{subsec:vexc_desired_ramp}, for a desired excess velocity, $v_{exc}$, we find an analytic expression for the corresponding apex radius which achieves that velocity at release. We then apply this formula to determine the minimum elevator length for which a Hohmann transfer from the Earth to each planet is possible using only excess velocity after release. Finally, we examine launch windows for non-Hohmann transfers by varying the start radius.
%

\subsection{Excess Velocity for Free Release from Apex Ramp}\label{subsec:vexc_notheta}
\begin{figure}
  \centering
  \includegraphics[width=.5\textwidth]{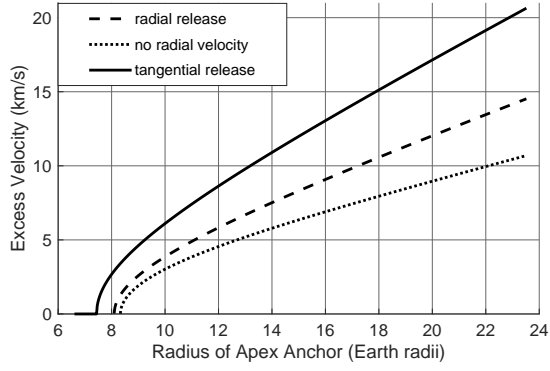}
  \caption{Excess velocity for tangential release (Tier 2, Eqn.~\eqref{eqn:vexc}), radial release (Tier 1, Eqn.~\eqref{eqn:T1_vexc_rad_max}), and release with no centripetal acceleration (Tier 0, Eqn.~\eqref{eqn:vexc_norad}) with release at Apex Anchor as a function of radial distance to apex anchor (subtract 1 ER for space elevator length) with $v_0=0$.}\label{fig:vexc}
\end{figure}

To determine excess velocity at release, we briefly revisit the total energy equations introduced in Subsection~\ref{subsec:T1_vr_vexc}.  To find the excess velocity, we again use the relationship $v_{exc}=\sqrt{2V}$. Recall that for $v_0=0$ and $r_0=r_g$, Equation~\eqref{eqn:SE_rocket} gives the maximum radial velocity as
\[
v_{r,\max}=\sqrt{\frac{2\mu}{r_p}+\omega_e^2 r_p^2 -3(\omega_e\mu)^{2/3}}.
\]
For a free release from an apex ramp with $\theta_c=90^\circ$, radial velocity is added to velocity of the apex anchor so that the magnitude of the velocity at release is given by
\begin{align*}
\norm{\vec v_{0-,T2}}^2&=(v_t+v_r)^2=(\omega_e r_p+v_r)^2\\
&=\omega_e^2 r_p^2+ \frac{2\mu}{r_p}+\omega_e^2 r_p^2 -3(\omega_e\mu)^{2/3}+2\omega_e r_p\sqrt{ \frac{2\mu}{r_p}+\omega_e^2 r_p^2 -3(\omega_e\mu)^{2/3}}.
\end{align*}
The expression for total energy is now
\begin{align*}
2V&=-\frac{2\mu}{r_p}+\norm{v}^2\\
&=-\frac{2\mu}{r_p} + \frac{2\mu}{r_p}+2\omega_e^2 r_p^2 -3(\omega_e\mu)^{2/3}+2\omega_e r_p\sqrt{ \frac{2\mu}{r_p}+\omega_e^2 r_p^2 -3(\omega_e\mu)^{2/3}}\\
&=2\omega_e^2 r_p^2 -3(\omega_e\mu)^{2/3}+2\omega_e r_p\sqrt{\frac{2\mu}{r_p}+\omega_e^2 r_p^2 -3(\omega_e\mu)^{2/3}}.
\end{align*}
Hence
\begin{equation}\label{eqn:vexc}
v_{exc,\max}=\sqrt{2\omega_e^2 r_p^2 -3(\omega_e\mu)^{2/3}+2\omega_e r_p\sqrt{\frac{2\mu}{r_p}+\omega_e^2 r_p^2 -3(\omega_e\mu)^{2/3}}}.
\end{equation}
This maximum excess velocity magnitude is shown in Figure~\ref{fig:vexc} and compared to the case of a non-sliding spacecraft (Tier 0, Eqn.~\eqref{eqn:vexc_norad}) and a sliding spacecraft with no apex ramp (Tier 1, Eqn.~\eqref{eqn:T1_vexc_rad_max}). This figure shows that the increase in excess velocity produced by sliding the spacecraft, while significant, is further amplified by the use of an apex ramp in Tier 2 space elevators.

In the case where a smaller excess velocity vector is desired, we may use
\begin{equation}\label{eqn:vexcmag_ramp_nomax}
v_{exc}=\sqrt{\norm{v}^2-\frac{2\mu}{r_p}}=\sqrt{(\omega_e r_p+v_r)^2-\frac{2\mu}{r_p}}
\end{equation}
where, for a desired $v_r$, the associated start radius is given by Equation~\eqref{eqn:rd_noramp}.

\subsection{Computing Turning Angle and Release Time for Free Transfer to Ecliptic}\label{subsec:TA1}
The analysis in this subsection parallels that in Subsection~\ref{subsec:T1_releasetime}, but is simpler due to the fact that the velocity vector at release is perpendicular to the radius vector. Specifically, from Section~\ref{sec:velocity}, we have the velocity vector in GCRF coordinates at exit from Earth's SOI.
\begin{align*}
[\vec v_{0+,T2}]_{GCRF}
&=\eta R_1(-\epsilon)R_3(\theta_{LST}+\theta_{TA})\bmat{0\\ \omega_e r_p+v_r\\0}\notag \\
&=R_1(-\epsilon)R_3(\theta_{LST}+\theta_{TA})\bmat{0\\ v_{exc}\\0}.
\end{align*}
where $\eta=\frac{v_{exc}}{v_r+\omega_e r_p}$ and $v_{exc}$ is given by Equation~\eqref{eqn:vexcmag_ramp_nomax}. This implies that a necessary and sufficient condition for $[\vec v_{0+,T2}]_{GCRF}\cdot \hat z=0$ in the case of free release from a Tier 2 elevator is
\[
\theta_{LST}+\theta_{TA}=90^\circ \quad \text{or}\quad \theta_{LST}+\theta_{TA}=270^\circ,
\]
which is equivalent to
\begin{equation}\label{eqn:thLST_ramp}
\theta_{LST}=90^\circ-\theta_{TA} \quad \text{or}\quad \theta_{LST}=270^\circ-\theta_{TA}.
\end{equation}
Again, the corresponding velocity vectors in GCRF at exit from the Earth's SOI are permanently fixed at
\begin{equation}
[\vec v_{0+,T2}]_{GCRF}=\bmat{\pm \sqrt{v_r^2+2v_r\omega_er_p+\omega_e^2r_p^2-2 \frac{\mu}{r_p} }\\0\\0}=\bmat{\pm v_{exc}\\0\\0}.
\end{equation}

Thus we again conclude that there are exactly two release times per day for which the velocity vector at exit from the Earth SOI lies in the ecliptic plane. To more precisely specify these times, we now compute the turning angle for the departure hyperbola from Earth's SOI.

\paragraph{Computing Turning Angle, $\theta_{TA}$}
To compute the turning angle, assuming $\theta_c=90^{\circ}$, we need to find the eccentricity of the departure hyperbola after release.
In this case, the position and velocity vectors at release are
\[
[\vec v_{0+}]_{SEI}=\bmat{0\\v_r+\omega_e r\\0}\qquad \text{and}\qquad [\vec r_{0+}]_{SEI}=\bmat{r_p\\0\\0}.
\]
The angular momentum vector is thus
\[
\vec h =[\vec r_{0+}]_{SEI} \times [\vec v_{0+}]_{SEI}=\bmat{0\\0\\(\omega_e r_p+v_r)r_p}
\]
which yields the eccentricity vector as
\begin{align*}
\vec e&=\frac{1}{\mu}\vec v \times \vec h-\frac{\vec r}{\norm{\vec r}}\\
&=\frac{1}{\mu}\bmat{0\\v_r+\omega_e r\\0} \times \bmat{0\\0\\(\omega_e r_p+v_r)r_p}-\bmat{1\\0\\0}=\bmat{\frac{1}{\mu}(v_r+\omega_e r)(\omega_e r_p+v_r)r_p-1\\0\\0}.
\end{align*}
The eccentricity is the magnitude of the eccentricity vector which is clearly
\[
e=\frac{1}{\mu}(v_r+\omega_e r)(\omega_e r_p+v_r)r_p-1=\frac{(v_r+\omega_e r)(\omega_e r_p+v_r)r_p-\mu}{\mu},
\]
with the resulting turning angle being
\[
\theta_{TA}=\sin^{-1}\left(\frac{1}{e}\right)=\sin^{-1}\left(\frac{\mu}{(v_r+\omega_e r)(\omega_e r_p+v_r)r_p-\mu}\right).
\]
In the case where we want to maximize excess velocity, $v_r$ is given by Equation~\eqref{eqn:SE_rocket} and we have
\[
e=1-\frac{r_p}{a}=1+\frac{r_p\left( 2\omega_e^2 r_p^3-3(\omega_e\mu)^{2/3}\right) +2\omega_e r_p^2\sqrt{ \frac{2\mu}{r_p}+\omega_e^2 r_p^2 -3(\omega_e\mu)^{2/3}}}{\mu}.
\]
Thus the turning angle gives us the 2 daily times ($\theta_{LST}$) for which we have free transfer to the ecliptic.

\subsection{Space Elevator Length as a Function of Desired Excess Velocity}\label{subsec:vexc_desired_ramp}
We now use Equation~\eqref{eqn:vexc} to find the apex anchor radius, $r_p$ which achieves a given desired maximum excess velocity, $v_d$, at exit from the Earth's SOI. This analysis parallels that of Subsection~\ref{subsec:T1_vexc_desired}. However, in the Tier 2 case, the expression for excess velocity in Eqn.~\eqref{eqn:vexc} in Subsection~\ref{subsec:vexc_notheta} is significantly more complicated. Specifically, we require
\[
v_d^2=2\omega_e^2 r_p^2 -3(\omega_e\mu)^{2/3}+2\omega_e r_p\sqrt{ \frac{2\mu}{r_p}+\omega_e^2 r_p^2 -3(\omega_e\mu)^{2/3}}
\]
or
\[
r_p^2 +r_p\sqrt{\frac{3(\omega_e\mu)^{2/3}}{\omega_e^2} + \frac{2\mu}{\omega_e^2 r_p}+r_p^2 }-\frac{v_d^2+3(\omega_e\mu)^{2/3}}{2 \omega_e^2}=0.
\]
This equation has the form
\[
r^2+r \sqrt{a+r^2+\frac{b}{r}}-c=0.
\]
where
\[
c=\frac{v_d^2+3(\omega_e\mu)^{2/3}}{2 \omega_e^2}, \qquad a=\frac{3(\omega_e\mu)^{2/3}}{\omega_e^2}, \qquad b=\frac{2\mu}{\omega_e^2}.
\]
Fortunately, there is an analytic solution to equations of this form. This solution is given by
\[
r_p=\frac{1}{2(a+2c)}\left(-b + \sqrt{b^2+4a c^2+8c^3 }\right).
\]
If we note that
\[
a+2c=\frac{v_d^2+3(\omega_e\mu)^{2/3}-3(\omega_e\mu)^{2/3}}{ \omega_e^2}=\frac{v_d^2}{\omega_e^2},
\]
then we get the slightly simpler expression:
\begin{equation}\label{eqn:vdes}
r_p=-\frac{2\mu}{v_d^2} + \frac{\omega_e^2}{v_d^2} \sqrt{b^2+4a c^2+8c^3 }.
\end{equation}
This solution yields the apex radius for which the maximum excess velocity at exit from Earth's SOI is $v_d$. The corresponding space elevator length is obtained by subtracting one Earth radius.
\subsection{Minimum Elevator Length for Hohmann Transfer to Distant Planets}
In this subsection, we repeat the analysis of Subsection~\ref{subsec:T1_hohmann} when an apex ramp is available by applying Equation~\eqref{eqn:vdes} to the excess velocity required for Hohmann transfer to each planet minus the velocity of the Earth. The results are listed in Table~\ref{tab:escape} and compared to the case of a Tier 1 elevator.
\begin{table}
  \centering
\begin{tabular}{ l | c| c}
  Destination & Tier 2 Length (km) & Tier 1 Length (km)  \\ \hline
  Venus & 44,195 & 50,650 \\
  Mars & 45,370 & 52,631 \\
  Jupiter & 71,030 & 93,299 \\
  Saturn & 79,432 & 106,036 \\
  Uranus & 85,135 & 114,596 \\
  Neptune & 87,327 & 117,870 \\
  Pluto & 88,268 & 119,274 \\
  Solar Escape & 91,398 & 123,919 \\
  Solar Escape + 10km/s & 154,687 & 216,288 \\
\end{tabular}
    \caption{Shortest Tier 1 and Tier 2 space elevators for free release Hohmann transfer to distant planets. Lengths for escape from the solar system, and escape from the solar system with 10km/s of excess velocity are also included.}\label{tab:escape}
\end{table}

Note that the required lengths for Tier 2 space elevators are significantly shorter than those for Tier 1. The cost of construction of a ramp may therefore be less than the cost of extending the elevator. Note, however, that the motivation for building a ramp is not necessarily the ability to deliver larger excess velocities, but rather the ability to rotate the ramp to match the ecliptic, enabling daily free transfers to the outer planets, as discussed in Sections~\ref{sec:ecliptic} and~\ref{sec:lambert} for the Tier 3 design. For Tier 2 elevators, we are still limited to departures at or near the winter or summer solstices and the associated requirements for planetary alignment.

\subsection{Launch Windows for non-Hohmann Transfers}
\begin{figure}[ht]
    \begin{subfigure}[t]{\textwidth}
  \centering
  \includegraphics[width=.75\textwidth]{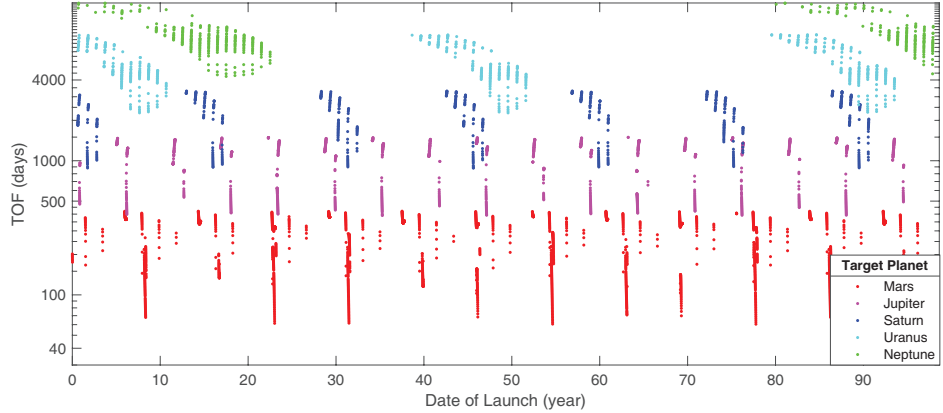}
  \caption{Times of Flight and launch dates for an apex radius of 100,000km.}\label{fig:TOFs_norotation_100000}
    \end{subfigure}
    \newline
    \begin{subfigure}[t]{\textwidth}
  \centering
  \includegraphics[width=.75\textwidth]{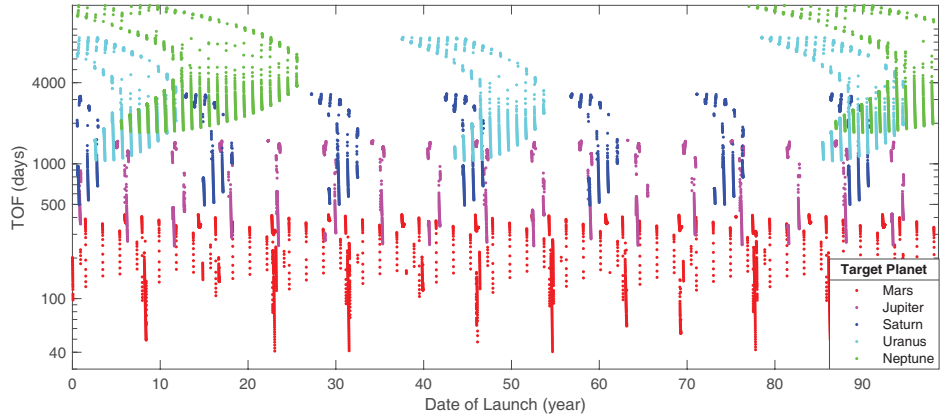}
  \caption{Times of Flight and launch dates for an apex radius of 150,000km.}\label{fig:TOFs_norotation_150000}
    \end{subfigure}
    \caption{Times of Flight and launch dates to for free release transfer to the outer planets for a Tier 2 space elevator over a 100 year interval. Each dot indicates a date on which a free release transfer is possible.}\label{fig:TOFs_norotation} \vspace{-5mm}
\end{figure}

In this subsection, we again consider using our ability to use start radius, $r_0$, to control radial velocity, $v_r$, which allows us to find non-Hohmann transfers to the outer planets. However, the results for Tier 2 elevators are not significantly different than those for Tier 1 elevators. The only difference is that the maximum excess velocity magnitude is increased. The results for Tier 2 elevators of apex radius 100,000km and 150,000km are given in Figures~\ref{fig:TOFs_norotation_100000} and~\ref{fig:TOFs_norotation_150000}, respectively. These figures show that the use of an apex ramp dramatically increases the utility of the 100,00km radius elevator, allowing us to reach all the outer planets. For the case of an apex radius of $r_p=150,000km$, the ramp significantly decreases the flight times to these outer planets. Note, however, that the ramp does not significantly decrease the time between launch windows over the Tier 1 elevator. For example, there is still a 65 year gap between free transfers to Neptune.

As mentioned previously, the motivation for including a ramp is not necessarily increased velocities, but also the ability to rotate the apex ramp to track the ecliptic, thereby eliminating the need to wait for the summer and winter solstices for interplanetary launch. Indeed, as will be shown in the following sections, rotation of the apex ramp allows for free release transfers to each of the outer planets every day of the year.
%
%

\section{Tier 3 Space Elevator: Transfer to Ecliptic via Apex Rotation}\label{sec:ecliptic}
In Sections~\ref{sec:thc=0} and~\ref{sec:escape}, we showed that free transfer to the planets and beyond is possible for both Tier 1 and Tier 2 space elevators. However, such launch windows are unreliable. Hence, in this section, we add an additional degree of freedom, $\theta_r$, enabled by a Tier 3 elevator, and show that by slowly rotating the apex anchor to track the ecliptic, free release transfer to the planets is possible on any given day (although some times of year yield faster transfers than others). Furthermore, we propose an efficient and reliable Newton-Raphson iteration for computing this rotation angle and the associated excess velocity vector in GCRF coordinates. Later, in Section~\ref{sec:lambert}, we will apply the results of this section to determine the minimum TOF to each of the outer planets as a function of both the length of the space elevator and the time of year.

The first goal of this section, then, is to calculate the rotation angle, $\theta_r$, for which the resulting excess velocity vector at exit of the Earth's SOI lies in the ecliptic plane. To do this we need to have an analytic expression for the $\hat z$ component of velocity in GCRF at exit from Earth's SOI as a function of $\theta_r$. To start, recall that Equation~\eqref{eqn:vexc_GCRF} from Section~\ref{sec:velocity} gives the velocity vector at exit from Earth's SOI in the GCRF coordinate system.
\begin{equation}\label{eqn:vexc_GCRF2}
[\vec v_{0+,T3}]_{GCRF}=R_1(-\epsilon)[\vec v_{0+}]_{ECI}=R_1(-\epsilon)R_3(\theta_{LST}) R_{1}(\theta_v)R_{3}(\theta_{TA})\bmat{0\\v_{exc}\\0}
\end{equation}
where recall $\epsilon=23.4^\circ$, $\theta_{LST}$ is known,
\[
\theta_v=\tan^{-1}\frac{v_r \sin(\theta_r)}{\omega_e r_p +v_r , \cos(\theta_r)},
\]
\[
v_{exc}=\sqrt{2V}=\sqrt{v_p^2-\frac{2\mu}{r_p}},
\]
\begin{equation}\label{eqn:vp_rotation}
v_p=\norm{[\vec v_{0-,T3}]_{SEI}}=\sqrt{\omega_e^2r_p^2+2\omega_e r_p v_r \cos(\theta_r)+v_r^2},
\end{equation}
and $v_r$ is the radial velocity, as determined by Equation~\eqref{eqn:rd_noramp} and which does not depend on $\theta_r$.
Thus we conclude that we have an expression for all parts of the velocity vector as a function of $\theta_r$ except for the turning angle, $\theta_{TA}$. We obtain such an expression in the following subsection.
\subsection{Orbital Elements and Turning Angle}
In this subsection, we need to obtain an expression for $\theta_{TA}$ as a function of $\theta_r$. This analysis is similar to that in Subsection~\ref{subsec:TA1} except for the need to account for the effect of rotation angle on the magnitude of velocity. Fortunately, because we assume $\theta_c=90^\circ$, periapse occurs at release and hence computation of the turning angle is only a function of the eccentricity,
\[
\theta_{TA}=\sin^{-1}\left(\frac{1}{e}\right).
\]
To compute eccentricity, there are several options. However, perhaps the simplest is to first compute the semimajor axis, $a$, from the expression for total energy,
\[
-\frac{\mu}{r_p}+\frac{v_p^2}{2}=-\frac{\mu}{2a}
\]
where $v_p$ is as given in Equation~\eqref{eqn:vp_rotation}. Therefore,
\[
a=\frac{1}{\frac{2}{r_p}-\frac{v_p^2}{\mu}}=\frac{r_p \mu}{2\mu-r_pv_p^2}.
\]
Now, since periapse is at the apex radius, we have $r_p=a(1-e)$. Thus
\[
e=1-\frac{r_p}{a}=1-r_p\left(\frac{2}{r_p}-\frac{v_p^2}{\mu}\right)=\frac{v_p^2 r_p}{\mu}-1.
\]
Applying the expression for $v_p$ in Equation~\eqref{eqn:vp_rotation}, we have that the turning angle can be computed from $\theta_r$ as
\[
\theta_{TA}=\sin^{-1}\left(\frac{\mu}{\omega_e^2 r_p^3+2\omega_e r_p^2 v_r \cos(\theta_r)+r_p v_r^2-\mu}\right).
\]

\subsection{Excess Velocity Component Normal to the Ecliptic}\label{subsec:zhat}
We may now compute $v_{z,GCRF}:=[\vec v_{0+,T3}]_{GCRF} \cdot \hat z$ as a function of $\theta_r$. Specifically, we expand Equation~\eqref{eqn:vexc_GCRF2} to obtain
\begin{align*}
v_{z,GCRF}=&v_{exc} \bbl(\sin(\epsilon) \sin(\theta_{LST}) \sin(\theta_{TA}) +
   \cos(\theta_{TA})\cos(\epsilon) \sin(\theta_{v})\\
   &- \cos(\theta_{TA})\cos(\theta_{LST}) \cos(\theta_{v}) \sin(\epsilon)\bbr).
\end{align*}
Now for insertion into the ecliptic we require $v_{z,GCRF}=0$, and hence we would like to choose $\theta_r$ such that
\begin{equation}\label{eqn:zhat}
\sin(\epsilon) \sin(\theta_{LST}) \sin(\theta_{TA}) =
   \cos(\theta_{TA}) \left( \cos(\theta_{LST}) \cos(\theta_{v}) \sin(\epsilon)-\cos(\epsilon) \sin(\theta_{v})\right).
\end{equation}

Before we proceed to the more general problem in the following subsection, there are two special cases of particular interest. In the first case, we choose $\theta_{LST}=90^\circ-\theta_{TA}$, and $\theta_r=0$. Then we have $\theta_v=0$ and Equation~\eqref{eqn:zhat} is satisfied. Thus even without rotation of the apex anchor, there are two daily free releases which transfer to the ecliptic plane - as discussed in Subsection~\ref{subsec:TA1}

In the second case, if we choose $\theta_{LST}=0$, then Equation~\eqref{eqn:zhat} is simply $\tan(\theta_v)=\tan(\epsilon)$ or $\theta_v=\epsilon$. Thus the solution is independent of turning angle and corresponds to the case where the departure hyperbola lies entirely in the ecliptic plane. Note, however, that we must still calculate $\theta_r$ from $\theta_v$ using the equation
\[
\sin(\theta_r)=\tan(\epsilon) \cos (\theta_r)+\frac{\omega_e r_p \tan (\epsilon)}{v_r}.
\]
The solution to this equation depends somewhat on $r_p$, and ranges between $0.87$ and $1.4$ radians - increasing as $r_p$ decreases.

In the following subsection, we propose a fast (typically only 3 iterations are required) Newton-Raphson iteration for computing $\theta_r$ in the more general case.
\subsection{Computing the Rotation angle, $\theta_r$, using Newton-Raphson Iteration}
There is no analytic expression for the roots of Equation~\eqref{eqn:zhat}. However, at least one real root always exists for $r_p>60579km$ and the equation is sufficiently simple that we may use Newton-Raphson iteration to find it. Newton-Raphson iterations are easy to implement, are known to converge very quickly (typically only requiring 3 iterations), and are the standard tool for solving Kepler's equation. For these reasons, we consider the existence of a Newton-Raphson solution to be relatively close to an analytic solution. In Section~\ref{sec:lambert}, we will consider the more difficult problem of TOF, for which there is no such algorithm.

For a given function $f(\theta_r)$ of a scalar variable ($\theta_r$) and with derivative $f'(\theta_r)$, the scaled ($\rho\le 1$) Newton-Raphson iteration is given as
\begin{equation}
\theta_{r,n+1}=\theta_{r,n}-\rho \frac{f(\theta_{r,n})}{f'(\theta_{r,n})}.\label{eqn:N-R}
\end{equation}
For Equation~\eqref{eqn:zhat}, we have (applying several inverse trig identities)
\begin{align*}
f(\theta_r)&=\tan(\epsilon) \sin(\theta_{LST}) \tan(\theta_{TA}) + \sin(\theta_{v})-\cos(\theta_{LST})\tan(\epsilon) \cos(\theta_{v})\\
&=\tan(\epsilon) \sin(\theta_{LST}) \frac{1}{\sqrt{e(\theta_r)^2-1}} + \frac{c_1(\theta_r)}{\sqrt{c_1(\theta_r)^2+c_2(\theta_r)^2}}\\
&\quad -\cos(\theta_{LST})\tan(\epsilon) \frac{c_2(\theta_r)}{\sqrt{c_2(\theta_r)^2+c_1(\theta_r)^2}},
\end{align*}
where
\begin{align*}
c_1(\theta_r)&=v_p \sin(\theta_r), \quad c_2(\theta_r)=\omega_e r_p +v_p \cos(\theta_r), \\
 e(\theta_r)&=\frac{\omega_e^2 r^3+2\omega_e r^2 v_r \cos(\theta_r)+r_p v_r^2-\mu}{\mu}.
\end{align*}

Now a simple application of the chain rule yields
\begin{align}
f'(\theta_r)=&\tan(\epsilon) \sin(\theta_{LST})\frac{d}{d\theta_r} \frac{1}{\sqrt{e(\theta_r)^2-1}} + \frac{d}{d\theta_r}\frac{c_1(\theta_r)}{\sqrt{c_1(\theta_r)^2+c_2(\theta_r)^2}}\notag \\
&-\cos(\theta_{LST})\tan(\epsilon) \frac{d}{d\theta_r}\frac{c_2(\theta_r)}{\sqrt{c_2(\theta_r)^2+c_1(\theta_r)^2}}.\label{eqn:fp}
\end{align}
where
\begin{align}
\frac{d}{d\theta_r}\frac{c_2(\theta_r)}{\sqrt{c_2(\theta_r)^2+c_1(\theta_r)^2}}
&=\frac{c_1(\theta_r)(c_1(\theta_r)c_2'(\theta_r)-c_2(\theta_r)c_1'(\theta_r))}{(c_2(\theta_r)^2+c_1(\theta_r)^2)^{3/2}}\\
\frac{d}{d\theta_r}\frac{c_1(\theta_r)}{\sqrt{c_2(\theta_r)^2+c_1(\theta_r)^2}}&=\frac{c_2(\theta_r)(c_2(\theta_r)c_1'(\theta_r)-c_1(\theta_r)c_2'(\theta_r))}{(c_2(\theta_r)^2+c_1(\theta_r)^2)^{3/2}}\\
\frac{d}{d\theta_r}\frac{1}{\sqrt{e(\theta_r)^2-1}}&=\frac{-e(\theta_r)e'(\theta_r)}{(e(\theta_r)^2-1)^{3/2}},\label{eqn:cprime}
\end{align}
and
\[ c_1'(\theta_r)  =v_r \cos(\theta_r),\quad c_2'(\theta_r)  =-v_r \sin(\theta_r),\quad   e'(\theta_r)  =-\frac{2\omega_e r_p^2 v_r \sin(\theta_r)}{\mu}. 
\]
Thus, combining the Newton-Raphson iteration in Equation~\eqref{eqn:N-R} with the formulae in Equations~\eqref{eqn:fp}-\eqref{eqn:cprime}, we have an efficient algorithm for finding $\theta_r$.

Because $f$ has continuous derivatives up to arbitrary order, the Newton-Raphson iteration has quadratic convergence in some neighborhood of the root. However, Newton-Raphson is not globally convergent, so we must ensure our initial guess is sufficiently close to the desired root. This is somewhat complicated by the fact that for large $r_p$ and most $\theta_{LST}$, $f(\theta)$ has two real roots - one where $v_r$ makes an acute angle with $v_t$ and one where the angle is obtuse. In the latter case, the solution may be discarded. Specifically, as indicated in Subsection~\ref{subsec:zhat}, the desired root (in radians) always lies in the interval $\theta_r \in[-1.4,1.4]$  - an interval which decreases as the radius of the apex anchor increases. To ensure we obtain the desired root, therefore, we restrict the range of $\theta_r$ so that the resulting eccentricity, $e>1$ -- a constraint which can be enforced using $\theta_r \in [-\theta_{r,\max},\theta_{r,\max}]$ where
\[
\theta_{r,\max}=\cos^{-1}\left(\frac{\mu}{2\omega_e r_p^2 v_r}-\frac{\omega_e^2r_p^3+rv_r^2-\mu}{2\omega_e r_p^2 v_r}\right).
\]
The algorithm is typically instantiated at $\theta_{r,0}=0^\circ$, as this is the point where $f'(\theta)$ is maximized in most cases. For all calculations in the following subsection, the algorithm converged to a tolerance of $10^{-6}$ within 10 iterations and never exceeded the interval $\theta_r \in [-\theta_{r,\max},\theta_{r,\max}]$. Furthermore, to improve convergence, if we are sweeping over several values of $\theta_{LST}$, it is better to use the previous solution to instance the subsequent value of $\theta_{r,0}$. Using this approach, only 4 iterations are required to obtain a tolerance of $10^{-6}$. Note that these limiting cases correspond to the case $r_p\cong60579km$ which is the minimum apex radius of a Tier 3 space elevator, above which a transition to the ecliptic may always be found. The corresponding minimum length is 54,201km. Note, however, that even at this minimum length, the excess velocity still has a minimum value of 3.8669 km/s - See Figure~\ref{fig:vexc_polar}. For elevators below this minimum, alternatives may be considered such as launch trajectories which pass out of the ecliptic.

\subsection{Excess Velocity Envelopes}
\begin{figure}
  \centering
  \includegraphics[width=.5\textwidth]{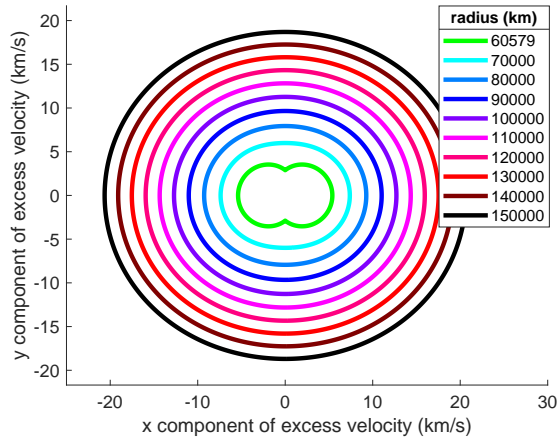}
  \caption{Excess velocity vector in GCRF coordinates after free release into ecliptic plane using ramp rotation in a Tier 3 elevator, parameterized by time of release and apex radius, $r_p$ (length of space elevator plus 6378km).}\label{fig:vexc_polar}
\end{figure}

In this subsection, we obtain data on the daily range of ecliptic-constrained excess velocity vectors in the GCRF coordinate system. To obtain this data, for a selection of space elevator lengths, we apply the Newton-Raphson iteration to every possible $\theta_{LST}$ in a sidereal day $\theta_{LST} \in [0,2\pi]$. Next, for each corresponding $\theta_r$, we plot $[\vec v_{0+}]_{GCRF}$, which gives the velocity vector in the ecliptic plane at exit from the Earth's SOI. This data is depicted in Figure~\ref{fig:vexc_polar} and is used in Section~\ref{sec:lambert} to compute the minimum Time of Flight (TOF) to each planet as a function of the time of year, as measured by the relative Earth ecliptic longitude.

\section{Tier 3 Space Elevator: Minimum Transfer Times to Planets}\label{sec:lambert}
\begin{figure}
\centering
\includegraphics[width=.35\textwidth]{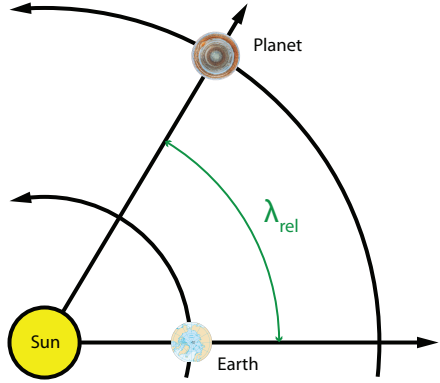}\vspace{-3mm}
\caption{Definition of relative ecliptic longitude, $\lambda_{rel}$. The period of the relative ecliptic longitude is the synodic period.}
\label{fig:ecl_long}
\end{figure}

Equipped with the results of Section~\ref{sec:ecliptic}, we may readily calculate the minimum-time ecliptic transfer to the each planet as a function of the apex radius and the time of year as measured by the difference in ecliptic longitude between the planet and the Earth (See Figure~\ref{fig:ecl_long}). The relative ecliptic longitude cycles from $0$ to $2 \pi$ radians over each synodic period. Note that while the synodic period for the outer planets is approximately 1 Earth year, the synodic period for Mars is 2.1 years and 1.6 years for Venus.

\subsection{Required Velocity Vector for given Transfer Time}
For a given transfer time (TOF), relative ecliptic longitude ($\lambda_{rel}$), and target planet, the required velocity vector at departure from the Earth SOI is given by the solution to Lambert's problem. Lambert's problem is a two-point boundary-valued problem. It takes two position vectors, $\vec r_1$ and $\vec r_2$ and a Time of Flight (TOF), and determines the velocity vectors, $\vec v_1$ and $\vec v_2$ corresponding to an elliptic or hyperbolic orbit which connects $\vec r_1$ and $\vec r_2$ with the resulting arc traversed in the specified TOF. Specifically, in this case, we take
\begin{equation}
\vec r_1=\bmat{d_e\\0\\0} \quad\text{and}\quad  \vec r_2=R_3(\lambda_{rel}+TOF \cdot n_{p})\bmat{d_p\\0\\0}\label{eqn:r1r2}
\end{equation}
where $d_e$ is the distance from the Sun to the Earth, $d_p$ is the distance from the Sun to the target planet, and $n_p$ is the mean motion of the target planet, given by
\[
n_p=\sqrt{\frac{\mu_s}{d_p^3}}
\]
where $\mu_s$ is the gravitational constant of the Sun.

The solution to Lambert's problem can be can be obtained by finding the solution to Lambert's equation. Unfortunately, the solution to this equation is not as amenable to Newton-Raphson iteration~\cite{avanzini_2008} as Kepler's equation or Equation~\eqref{eqn:zhat}. One part of the problem is that the transition from elliptic to hyperbolic orbits results in a change to the equation. This problem can be partially resolved through the use of universal variables, as described in~\cite{lancaster_1969,vallado_book,battin_book}. However, this change of variables complicates standard root-finding algorithms such as Newton-Raphson. Thus, the most reliable method for solving Lambert's problem remains the use of bisection - an approach we take in this paper, using an implementation from~\cite{vallado_book}. The disadvantage of bisection is that the rate of convergence is rather poor. To accelerate the convergence, many alternative solutions to Lambert's problem have been proposed, such as in~\cite{avanzini_2008,ahn_2013,zhang_2010,gooding_1990,izzo_2015,prussing_2000}. Unfortunately, however, our experience has found that for hyperbolic orbtis and long times of flight, these implementations are difficult to reproduce or are unreliable.
%
\begin{figure}
\centering
\includegraphics[width=.5\textwidth]{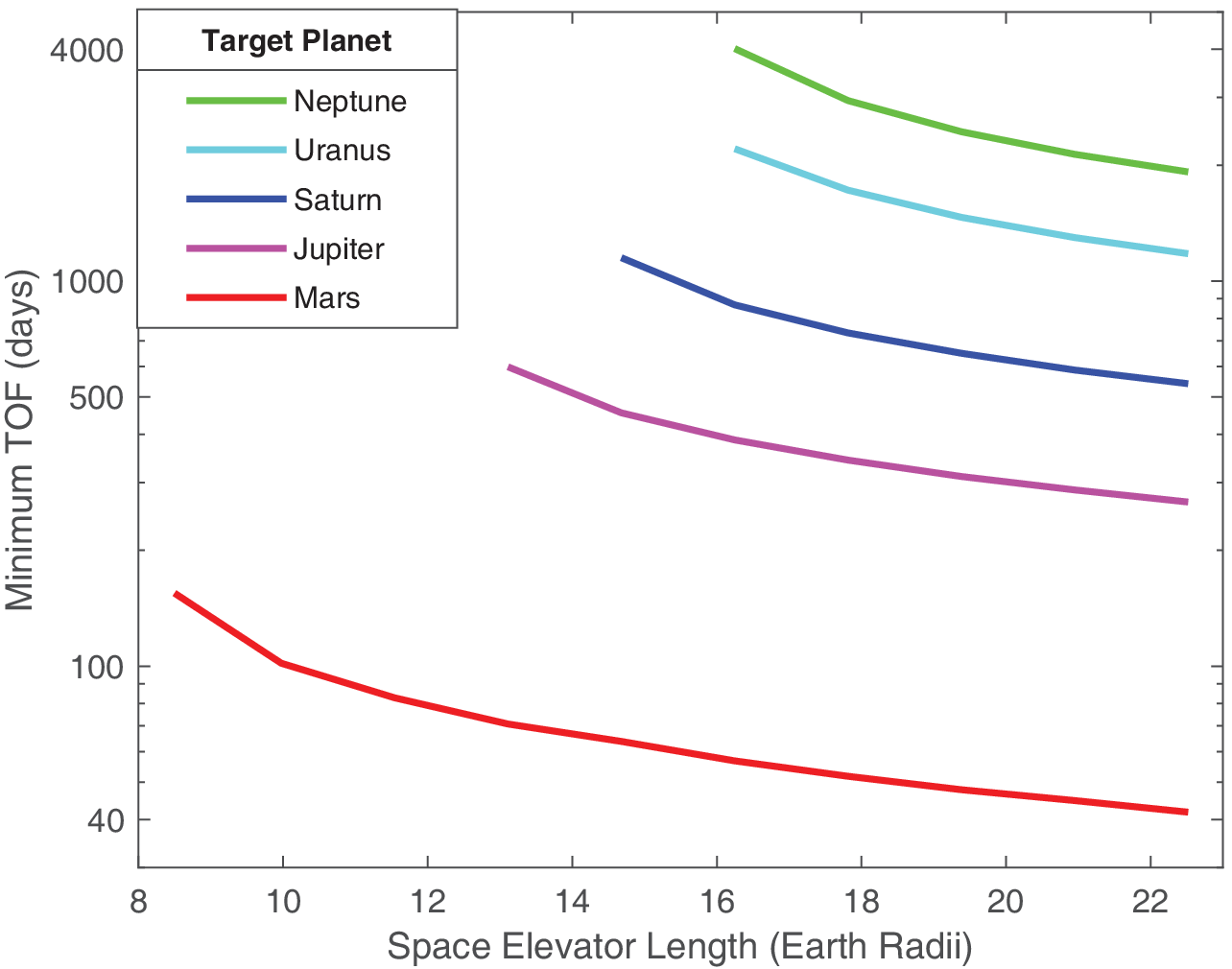}\vspace{-3mm}
\caption{Semilog plot of minimum yearly free release Times-of-Flight (TOF) in days from a Tier 3 space elevator to each of the outer planets as a function of length of the elevator. The Minimum TOF to each planet is available once per synodic period.}
\vspace{-5mm}\label{fig:TOF_mins}
\end{figure}

Thus, in the following subsection, for any  given $TOF$ and $\lambda_{rel}$, we use the universal variables implementation combined with bisection to compute the corresponding required barycentric (BCRS) velocity vector at departure from the Earth's SOI, $\vec v_1$. We then determine if $\vec v_1-\vec v_e$ ($\vec v_e$ is the velocity vector of the Earth) lies in the excess velocity envelope as calculated in Section~\ref{sec:ecliptic} and illustrated in Figure~\ref{fig:vexc_polar}.

\subsection{Computing Minimum TOF as a Function of Relative Ecliptic Longitude}
Given a solution to Lambert's problem, calculating the minimum daily TOF is relatively straightforward, if inelegant. For each day in the synodic period, as measured by relative ecliptic longitude (See Figure~\ref{fig:ecl_long}), we test feasibility of each TOF, incrementing by one day until a feasible TOF is found. The minimum TOF used to initialize the sweep is determined by $TOF_{\min}=\frac{\norm{d_p-d_e}}{v_{exc}+v_e}$, which corresponds to the smallest distance between the Earth and target planet divided by the velocity of the spacecraft at departure from Earth's SOI. Note this minimum is only valid for travel to outer planets.

To test feasibility of a TOF, we calculate the target position of the planet at end of the TOF, as determined by Equation~\eqref{eqn:r1r2}. This yields the associated velocity at departure, $\vec v_1$. We then determine whether $\vec v_1-\vec v_e$ lies in the excess velocity envelope using the data obtained in Section~\ref{sec:ecliptic} (Figure~\ref{fig:vexc_polar}). If not, we increase the time of flight by one day and repeat until a feasible TOF is found. While this solution is computationally inefficient, it is necessary as the required excess velocity is not a convex, monotone, or even smooth function of the TOF. Note also, we check both the direct and retrograde solutions to Lambert's problem in these calculations.
%
\begin{figure}
\centering
\includegraphics[width=.5\textwidth]{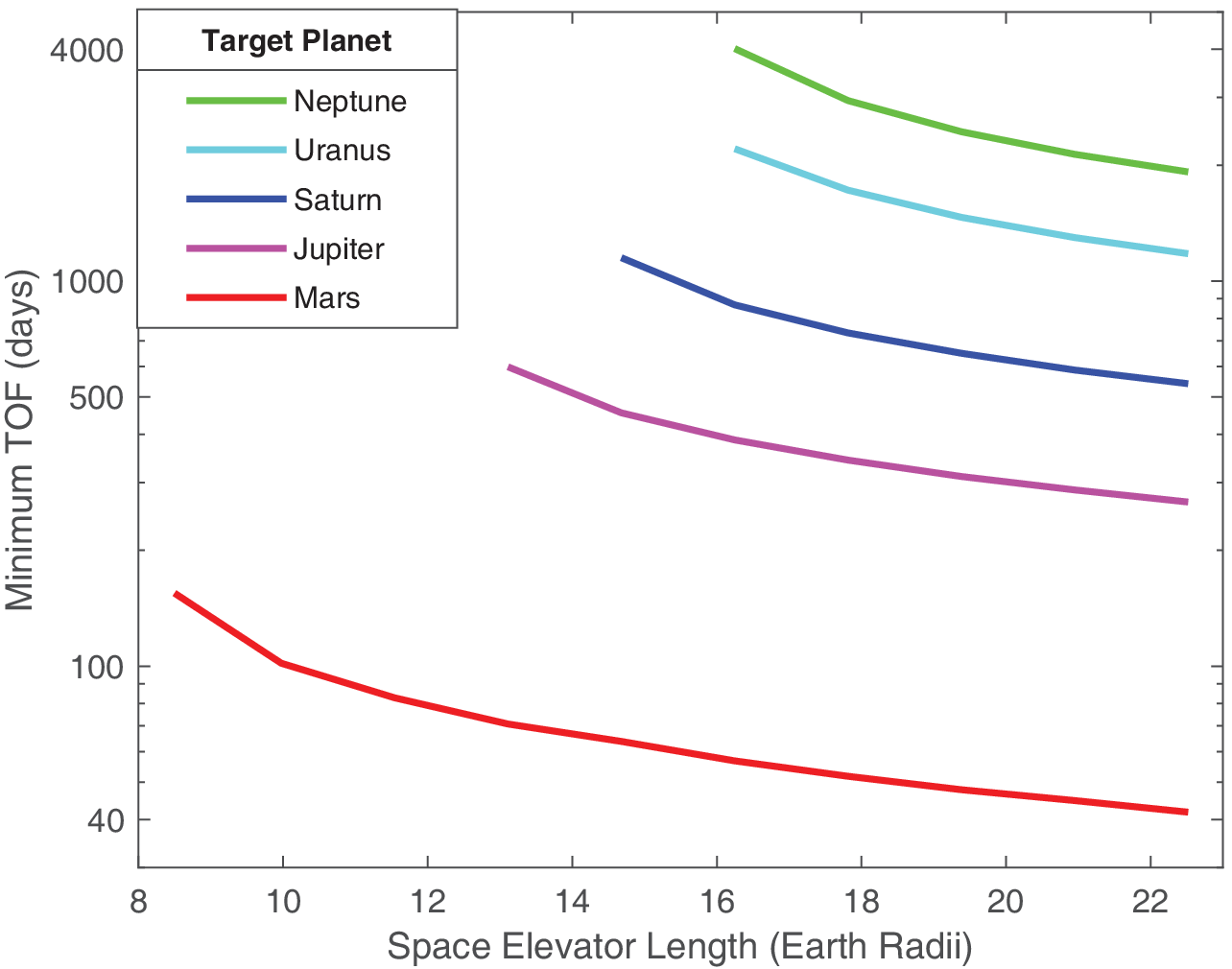}\vspace{-3mm}
\caption{Semilog plot of minimum yearly free release Times-of-Flight (TOF) in days from a Tier 3 space elevator to each of the outer planets as a function of length of the elevator. The Minimum TOF to each planet is available once per synodic period.}
\vspace{-5mm}\label{fig:TOF_mins}
\end{figure}

\begin{figure}[ht]
    \begin{subfigure}[t]{0.3\textwidth}
        \centering
\includegraphics[width=\linewidth]{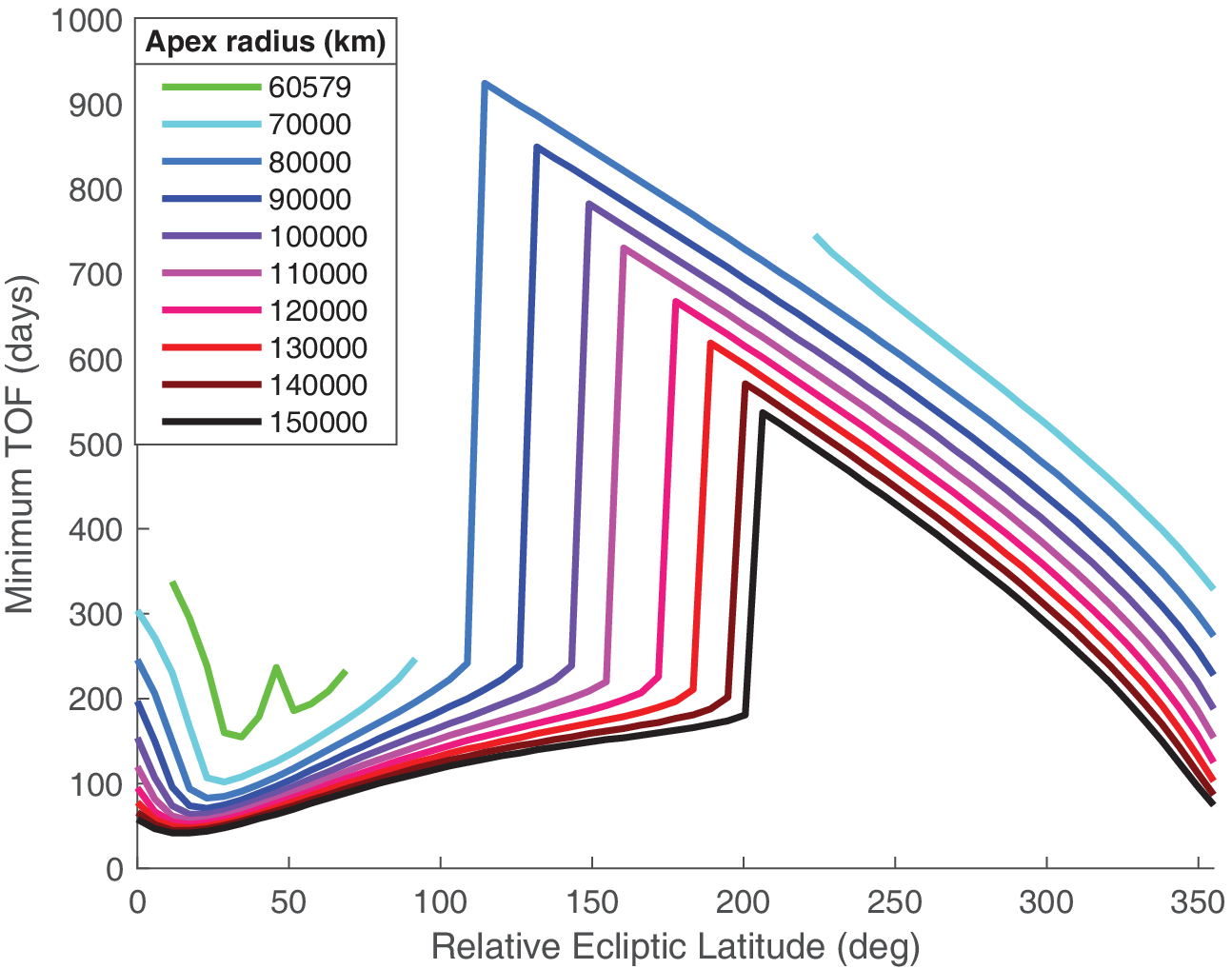}
        \caption{Times of Flight to Mars.}\label{subfig:TOF_mars}
    \end{subfigure}\hfil
    \begin{subfigure}[t]{0.3\textwidth}
        \centering
\includegraphics[width=\linewidth]{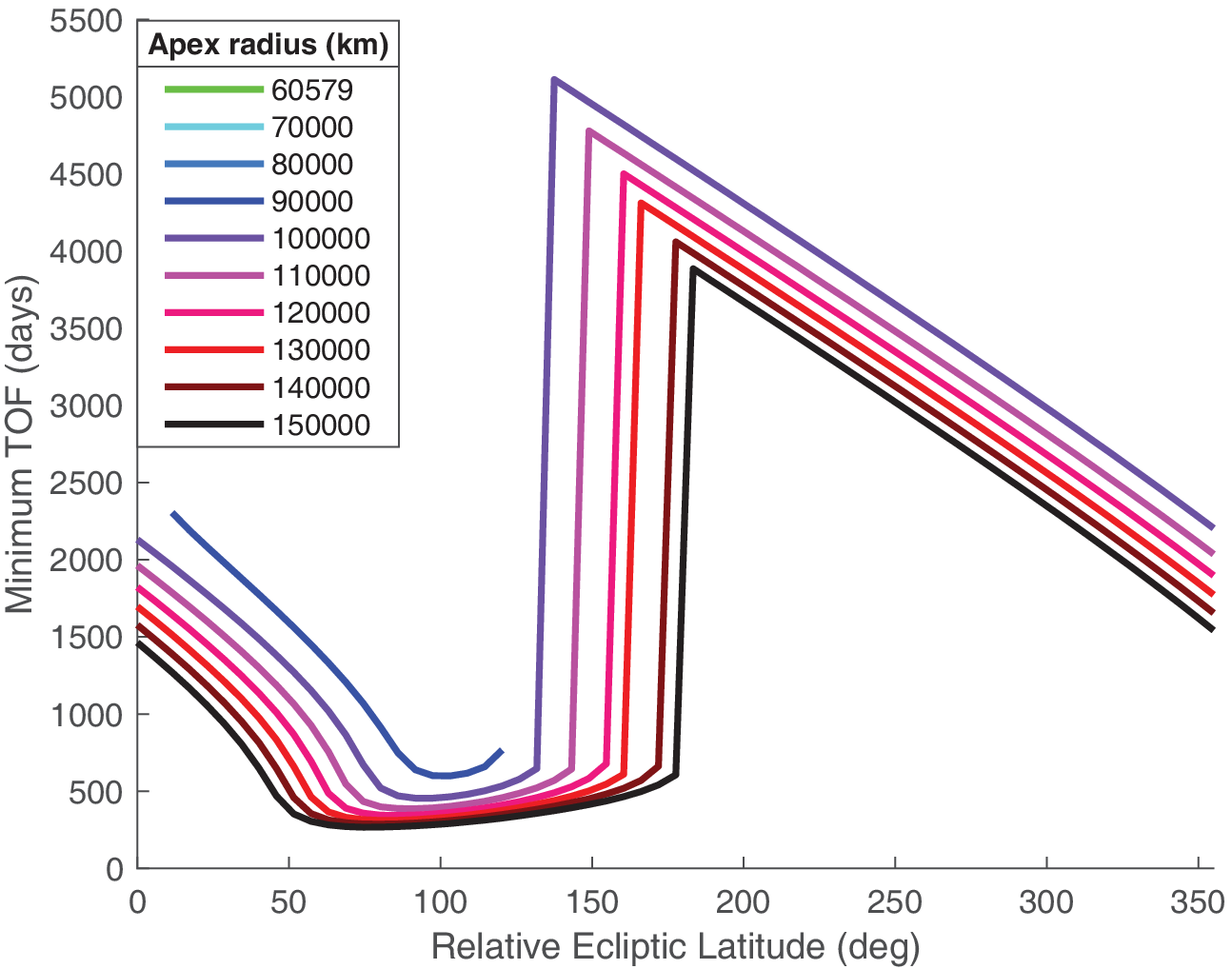}
        \caption{Times of Flight to Jupiter.}\label{subfig:TOF_jupiter}
    \end{subfigure}\hfil
    \begin{subfigure}[t]{0.3\textwidth}
        \centering
\includegraphics[width=\linewidth]{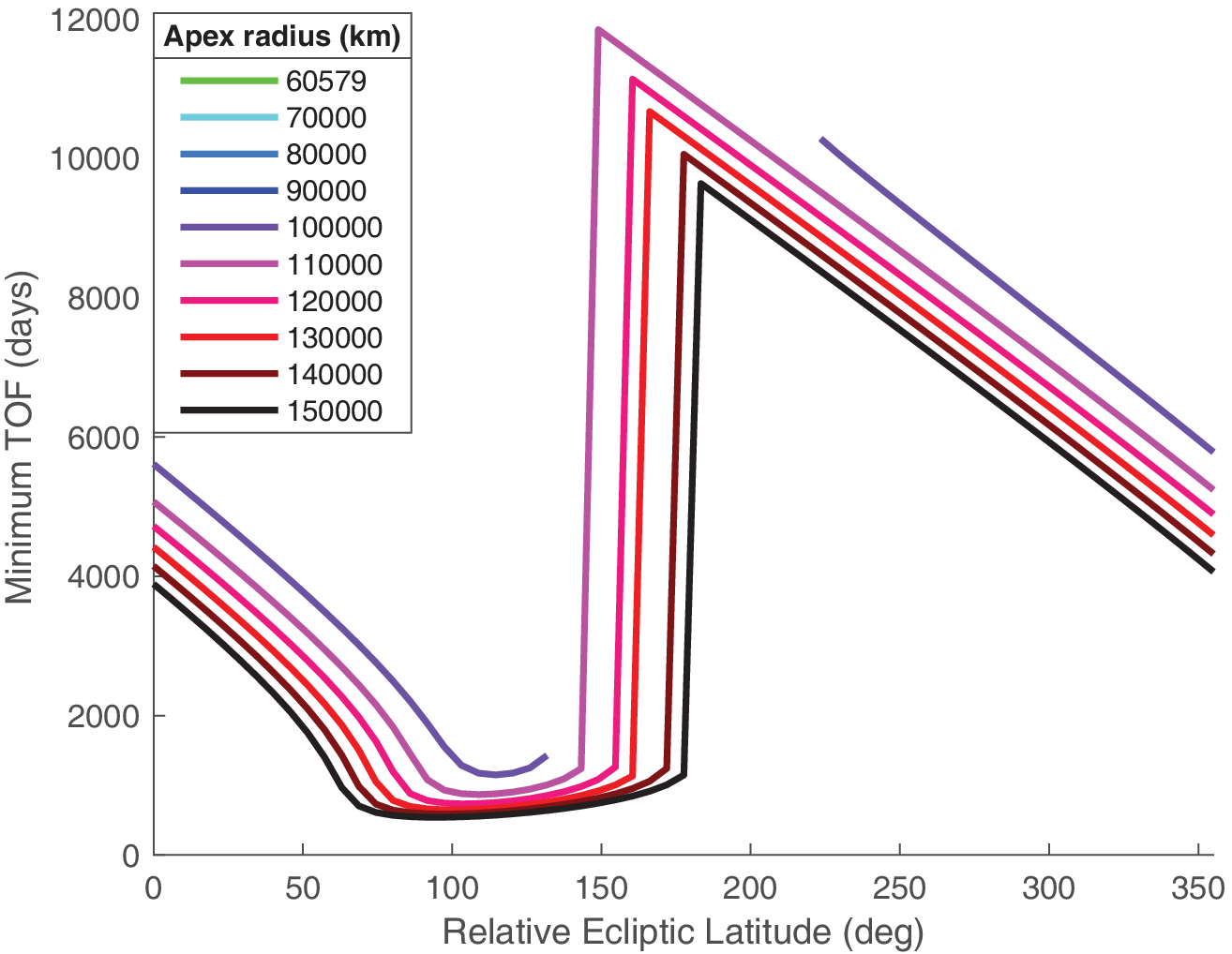}
        \caption{Times of Flight to Saturn.}\label{subfig:TOF_saturn}
    \end{subfigure}
    \newline
        \begin{subfigure}[t]{0.3\textwidth}
        \centering
\includegraphics[width=\linewidth]{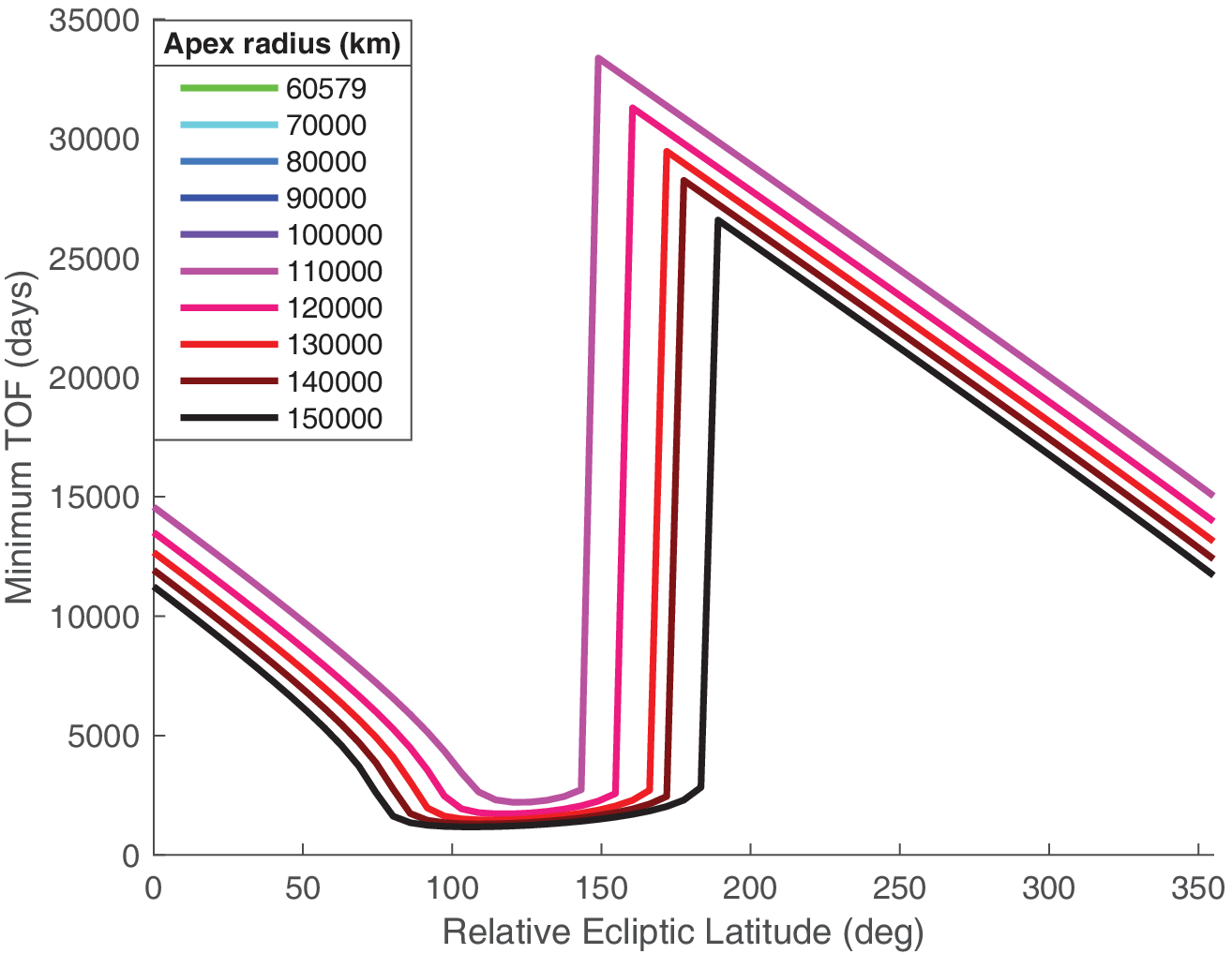}
        \caption{Times of Flight to Uranus.}\label{subfig:TOF_uranus}
    \end{subfigure}
        ~~
        \begin{subfigure}[t]{0.3\textwidth}
        \centering
\includegraphics[width=\linewidth]{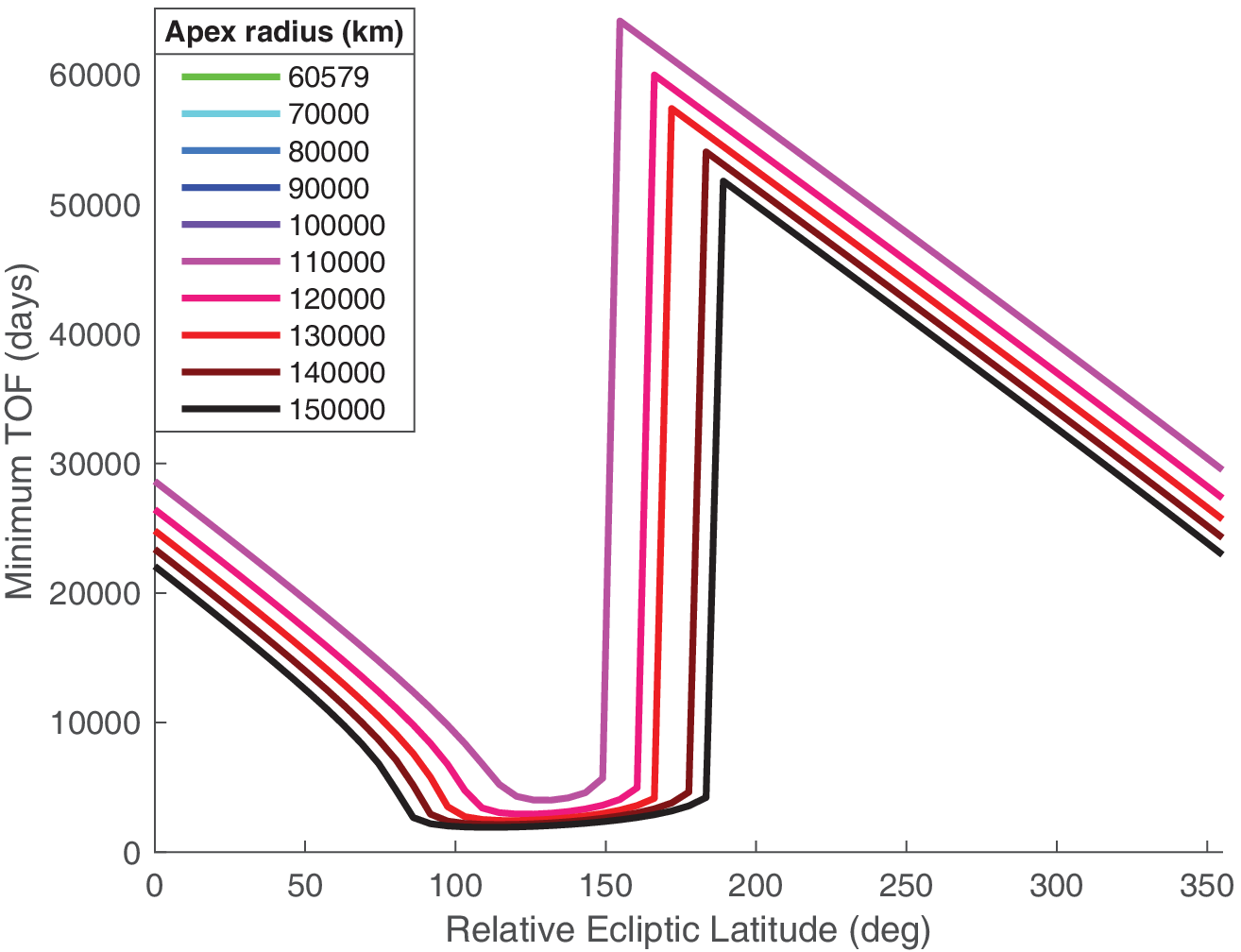}
        \caption{Times of Flight to Neptune.}\label{subfig:TOF_neptune}
    \end{subfigure}\hfill
    \caption{Minimum Times of Flight (TOF) to the outer planets as a function of launch date (as measured by relative ecliptic longitude) and radius of the apex anchor, $r_p$, over one synodic period. Minimum flight times are computed using the smallest TOF solution to Lambert's problem which lies in the excess velocity envelope depicted in Figure~\ref{fig:vexc_polar}. Note that for certain times of year, TOF can be reduced by waiting. However, use of this strategy is not considered in the Subfigures~\ref{subfig:TOF_mars}-\ref{subfig:TOF_neptune}.}\label{fig:TOFs} \vspace{-5mm}
\end{figure}

The results are illustrated in Figures~\ref{fig:TOFs} for each planet, listing the smallest time of flight as a function of the relative ecliptic longitude. Note that there are certain times-of-year when it is better to wait, as the TOF curve occasionally decreases faster than time elapses. Additionally, in Figure~\ref{fig:TOF_mins}, we list the minimum yearly TOF to each planet over one synodic period.

As a final note, it is possible to use the $\vec v_2$ values from Lambert's problem to determine the $\Delta v$ required for insertion into planetary orbit. As an alternative approach, one might minimize this $\Delta v$ at planetary insertion rather than TOF. However, we neglect this analysis for two reasons. The first is that the focus of this paper is on propellant-free transfers. The second is that there is no particular launch capacity of the space elevator in terms of propellant mass - hence the $\Delta v$ budget of the spacecraft does not directly affect the orbital mechanics of the transfer problem.

\section{Tier 3 Space Elevator with Slingshot Maneuvers}\label{sec:trebuchet}
While the times of flight described in the previous section are certainly shorter than existing options using rockets, even for Tier 3 elevators, the maximum excess velocity at departure from the Sun's SOI is still not realistic for interstellar exploration. Therefore, in this final section, we examine the use of a retrograde ramp on a Tier 3 space elevator for significantly increasing the velocity of spacecraft at release. As is the case throughout this paper, the only source of energy to be considered is the rotational energy of the Earth (no propellent). The underlying assumption, however, is that the space elevator is capable of launching high-mass spacecraft.

\begin{figure}[ht]
    \begin{subfigure}[t]{0.3\textwidth}
        \centering
\includegraphics[scale=1.3]{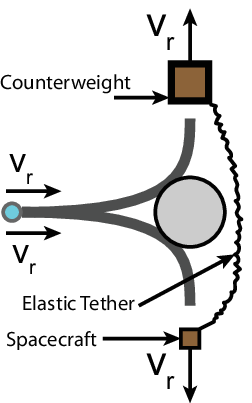}
        \caption{\textbf{[Step 1]} The spacecraft is launched using the retrograde ramp and the counterweight is simultaneously launched using the posigrade ramp. The relative velocity at release between spacecraft and counterweight is $2v_r$. Spacecraft and counterweight are connected by an elastic tether.}\label{subfig:step1}
    \end{subfigure}\hfil
    \begin{subfigure}[t]{0.3\textwidth}
        \centering
\includegraphics[scale=1.4]{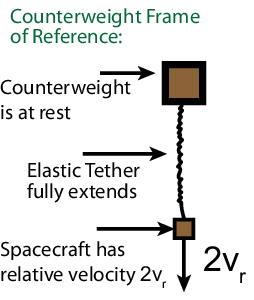}
        \caption{\textbf{[Step 2]} View from the reference frame of the counterweight after release from the space elevator. The spacecraft is moving at relative velocity $2v_r$. The tether fully extends and becomes taught. }\label{subfig:step2}
    \end{subfigure}\hfil
    \begin{subfigure}[t]{0.3\textwidth}
        \centering
\includegraphics[scale=1.4]{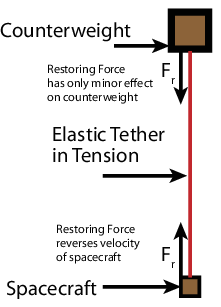}
        \caption{\textbf{[Step 3]} The extended elastic tether exerts equal and opposite reaction forces on the spacecraft and counterweight. By assumption, the mass of the counterweight is significantly greater than that of the spacecraft and hence the force $F_r$ causes minimal acceleration of the counterweight.}\label{subfig:step3}
    \end{subfigure}
    \newline
        \begin{subfigure}[t]{0.3\textwidth}
        \centering
\includegraphics[scale=1.4]{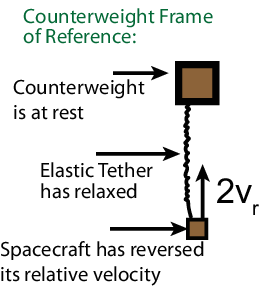}
        \caption{\textbf{[Step 4]} The elastic tether has effectively reversed the direction (but not the magnitude) of the velocity of the spacecraft in the frame of reference of the counterweight. The tether becomes slack and the spacecraft passes by the counterweight.}\label{subfig:step4}
    \end{subfigure}
        ~~
        \begin{subfigure}[t]{0.3\textwidth}
        \centering
\includegraphics[scale=1.4]{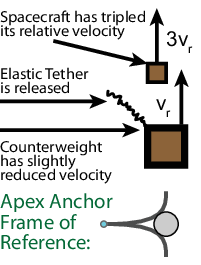}
        \caption{\textbf{[Step 5]} After the spacecraft passes the counterweight at velocity $2v_r$, the tether is released. The velocity of the spacecraft is now $3v_r$ with respect to the apex anchor. In the ECI frame, the velocity of the apex anchor is now added to that of the spacecraft.}\label{subfig:step5}
    \end{subfigure}\hfill
    \caption{An illustration of the 5 stages of the space elevator slingshot maneuver for a light spacecraft and heavy counterweight launched simultaneously using retrograde and prograde ramps, respectively. The final velocity of the light spacecraft in Step 5, Subfigure~\ref{subfig:step5} is $3v_r$ in the prograde direction relative to the apex anchor.}\label{fig:steps} \vspace{-5mm}
\end{figure}

For a Tier 3 space elevator slingshot maneuver, we consider use of the retrograde ramp, as depicted in Figure~\ref{fig:design}. The principles involved are similar to those used for slingshot maneuvers in planetary gravity assists. Specifically, both a light spacecraft and a heavy counterweight (analogous to a planet) are launched simultaneously in opposite directions, both with velocity $v_r$ with respect to the apex anchor (Subfigure~\ref{subfig:step1}). The counterweight is launched in the desired prograde direction of motion, while the spacecraft is launched in the opposite, retrograde direction. The relative velocity of counterweight and spacecraft is then $2v_r$, where the direction of the spacecraft in the inertial frame of the counterweight is retrograde (Subfigure~\ref{subfig:step2}). The counterweight and spacecraft are connected by an elastic tether (a spring), which reverses the direction of motion of the spacecraft in the inertial frame of the counterweight but does not significantly alter the velocity of the counterweight (Subfigure~\ref{subfig:step3}). The velocity of the spacecraft is now aligned with that of the counterweight (Subfigure~\ref{subfig:step4}). In the frame of the apex anchor, however, the velocity of the counterweight is now added to that of the spacecraft, to get a velocity of $3v_r$ with respect to the apex anchor (Subfigure~\ref{subfig:step5}). The velocity of the apex anchor is now added to that of the spacecraft to get a velocity in the inertial frame of magnitude $3v_r+v_t$. Recall that the maximum $v_r$ is as given by Equation~\eqref{eqn:SE_rocket}, is depicted in Figure~\ref{fig:vr}, and achieves $10km/s$ at a space elevator of length 22 Earth radii. Note that in this simplified analysis, if the spacecraft is capable of independently producing velocity change $\Delta v$, and does this before the tether is extended, then the final velocity magnitude is $3v_r+v_t + 2\Delta v$.

For the simplified analysis depicted in Figure~\ref{fig:steps}, however, we assumed the motion of the counterweight was not significantly affected by the velocity reversal of the spacecraft. However, in practice, the counterweight has finite mass, which we label as $m_2$ and is larger than the mass of the spacecraft, which we label $m_1$. In this case, we can calculate the actual velocity of the spacecraft relative to the apex anchor after the slingshot maneuver as
\[
v_{rel}=\frac{3m_2-m_1}{m_1+m_2}v_r.
\]
For example, if $m_2=10m_1$, then the velocity of the spacecraft relative to the apex anchor is reduced to $2.63 v_r$.

\paragraph{A Tier 3 Slingshot Staging Design} As a final note, the slingshot design may be extended to multiple counterweights by staging (similar to rocket staging). For example, consider launching a  first counterweight (very large) prograde, tethered to a second counterweight (large) launched retrograde, tethered to a spacecraft (small) launched retrograde. In this case, the second counterweight initially acts as the spacecraft and is accelerated to a prograde velocity $3 v_r $ with respect to the apex anchor by the first counterweight (ignoring the scaling factor introduced above). The velocity of the second counterweight with respect to the spacecraft is then $4v_r$. This second counterweight then reverses the relative motion of the spacecraft in what is now the prograde frame of the second counterweight. The velocity of the spacecraft with respect to the apex anchor is now $3v_r+4v_r=7v_r$. Indeed, if we define $n_{k}$ to be the $v_r$ multiplier for a $k$-stage slingshot, then $n_0=1$ and $n_k$ is given by the recurrence relation $n_k=2n_{k-1}+1$ - a geometric series! Thus the velocity magnitude relative to the apex anchor for a $k$-stage slingshot is
\[
v_k= (2^{k+1}-1) v_r
\]
The first several multiples in this sequence are $\{1,3,7,15,31\}$. Thus, with strong tethers, it is possible to achieve almost any desired velocity using a relatively small number of stages.

%
%
%
%

%
%

\section{Conclusion}
In this paper, we have presented the fundamental orbital mechanics of operation and utilization of a space elevator launch system. We have shown that space elevators, by extending the radius of the Earth past the point where centripetal acceleration is stronger than gravity, allow us to harness the unlimited energy of Earth's rotational inertial to launch spacecraft at velocities far beyond anything possible using current rocket technology.

Specifically, we have proposed 3 tiers of space elevator: elevators without an apex ramp; elevators with an apex ramp, but without rotation of the ramp; and elevators capable of slowly rotating an apex ramp. In all cases, we have shown that even moderately sized elevators can propel spacecraft to velocities beyond the escape velocity of the solar system - and without any use of rockets. However, for Tier 1 and 2 elevators, transfer to the ecliptic plane poses serious constraints on the available launch dates - making planetary transfer opportunities using free release infrequent, with gaps of up to 65 years between launch windows. The rotation of the apex ramp in Tier 3 elevators, by contrast, does not increase the magnitude of the spacecraft velocity, but rather allows for launch into the ecliptic plane without any constraint on launch time - resulting in planetary free release transfers to every planet every day of the year. However, we have also found that the times of flight for these daily launch opportunities vary throughout the year, with the shortest transfer to Mars being 40 days or 200 days to Neptune. Finally, we have shown that planetary gravity assist maneuvers have a space elevator equivalent in the form of slingshot maneuvers and these maneuvers can potentially be staged to achieve any desired launch velocity.

Throughout this paper, we have endeavored to find and apply new mathematical laws and relationships - minimizing or eliminating the use of numerical algorithms whenever possible. We hope that these formulae might prove the basis for improved understanding and appreciation for the elegance and power of space elevators as a tool for deep space exploration.

And so we bring an end to our story of space elevators - the great arms by which mankind might one day reach the heavenly bodies. It has been said that the book of nature is writ in the language of mathematics, and its characters are triangles, circles, and other geometrical figures, without which it is humanly impossible to understand a single word of it; without these, one is wandering around in a dark labyrinth~\cite{galilei_book}. If that is so, then perhaps, by continued and rigorous mathematical study, we might yet hope to shed a little light on the next chapter.

\section*{Acknowledgement} Thanks are due to Peter Swan and ISEC for an introduction to the unsolved problems in space elevator launch systems.

\bibliographystyle{plain}
\bibliography{peet_bib,space_elevator}

\end{document}